\documentclass[11pt,a4paper]{article}
\usepackage{amsmath}

\newtheorem{lemma}{Lemma}
\newtheorem{identity}[lemma]{Identity}
\newtheorem{conjecture}[lemma]{Conjecture}
\numberwithin{equation}{section}
\numberwithin{lemma}{section}
\begin{document}

\title{\bf{Autocorrelation of Random Matrix Polynomials }}
\vspace {2 in}

\author{J.\ B.\ Conrey$^{1}$, D.\ W.\ Farmer$^{1}$, J.\ P.\ Keating$^{2}$, M.\ O.\ Rubinstein$^{1}$
 and N.\ C.\ Snaith$^{2}$\\
        1. American Institute of Mathematics,\\
        360 Portage Ave,\\
        Palo Alto, CA 94306, USA\\
        and\\
        2. School of Mathematics,\\ University of Bristol,\\
            Bristol BS8 1TW, UK}
\date{\today}
\maketitle
\thispagestyle{empty}
\vspace{.5cm}
\begin{abstract}
 We calculate the autocorrelation functions (or shifted moments) of the
 characteristic polynomials of matrices drawn uniformly with respect to
 Haar measure from
 the groups $U(N)$, $O(2N)$ and $USp(2N)$.  In each case the result can be
expressed in three equivalent forms: as a determinant sum (and hence in terms of symmetric
 polynomials), as a combinatorial sum, and
as a multiple contour integral.  These formulae are analogous to
those previously obtained for the Gaussian ensembles of Random
Matrix Theory, but in this case are identities for any size of
matrix, rather than large-matrix asymptotic approximations.   They
also mirror exactly the autocorrelation formulae conjectured to
hold for $L$-functions in a companion paper. This then provides
further evidence in support of the connection between Random
Matrix Theory and the theory of $L$-functions.
\end{abstract}

\section{Introduction}
The conjectured connection between random matrices and number
theory dates back to an exchange between H.~L.~Montgomery and
F.~J.~Dyson \cite{kn:mont73} in which they discovered that the
two-point correlation function of the zeros of the Riemann zeta
function, studied by the former, is the same, in the appropriate
limit, as the two-point correlation function of the eigenvalues of
random matrices, calculated by the latter.  Since then
calculations of the three-point zero correlation function by
Hejhal \cite{kn:hejhal94}, the general $n$-point zero correlation
functions by Rudnick and Sarnak \cite{kn:rudsar} and Bogomolny and
Keating \cite{kn:bogkea95,kn:bogkea96}, the study of the low-lying
zeros of families of $L$-functions by Katz and Sarnak
\cite{kn:katzsarnak99a}, and extensive numerical computations
\cite{kn:odlyzko89,kn:rub98} have strengthened the connection.

In the past few years, following the work of Keating
and Snaith \cite{kn:keasna00a,kn:keasna00b}, Conrey and Farmer
\cite{kn:confar00} and Hughes,
Keating and O'Connell \cite{kn:hughes00,kn:hughes01}, it has
become clear that the leading order asymptotics of the
mean values (or moments) of the Riemann zeta function and
families of $L$-functions can be understood, again conjecturally,
in terms of
the corresponding value distribution of the characteristic polynomials of
random
matrices. In the random matrix case, the average is performed with respect
to Haar measure for either the group of unitary ($U(N)$), orthogonal
($O(2N)$) or
unitary symplectic ($USp(2N)$) matrices, depending on the symmetries of
the family in question.

Our purpose here is to calculate the autocorrelation functions (sometimes
called the shifted moments) for the
characteristic polynomials of random matrices from the groups just listed.
Specifically, let $\Lambda_M(s)$ represent the characteristic polynomial of a matrix
$M$ associated with an element of a compact group $G$, and let $dM$ denote Haar measure on
$G$.  We calculate
\begin{equation}
\int_{G}\Lambda_M(s_1^{-1})\cdots
\Lambda_M(s_m^{-1})\Lambda_{M^{\dagger}}(s_{m+1})\cdots
\Lambda_{M^{\dagger}}(s_{n}) dM
\label{eq:intro1}
\end{equation}
when $G=U(N)$ (here $M^{\dagger}$ is the Hermitian
conjugate of $M$), and
\begin{equation}
\int_G \Lambda_M(s_1^{-1})\cdots \Lambda_M(s_k^{-1}) dM
\label{eq:intro2}
\end{equation}
when $G=O(2N)$ and $G=USp(2N)$.  (The reason for having a different
definition in the first case is related to symmetries in the eigenvalue spectra.)  In each case the
result will be presented in three equivalent forms: as a determinant sum, in the
 style of Basor and Forrester \cite{kn:basfor94} (and hence in terms of symmetric
 polynomials); as a combinatorial sum; and
as a contour integral, in the style of Br\'ezin and Hikami
 \cite{kn:brezhik00}.

Conjectures based on these random matrix results for the
autocorrelation functions of $L$-functions are presented in a
companion paper to this one \cite{kn:cfkrs}.  We here prove the
results stated there.  The combination of the random matrix
results derived here and the numerical evidence in favour of the
conjectures for $L$-functions put forward in \cite{kn:cfkrs} add
considerable weight to the idea that there are fundamental
connections between the two subjects. In addition, the random
matrix calculations carry an interest of their own in connection
with work on Toeplitz matrices {\cite{kn:basfor94,kn:bumdia02} in
the unitary case, and with the elegant dual pair method of
Zirnbauer and Nonnenmacher \cite{kn:nonzir02} for all three of the
above mentioned compact groups.

Similar calculations to those described here have been performed
on ensembles of Hermitian matrices, first by Andreev and Simons
\cite{kn:andsim95} and then by Br\'ezin and Hikami
\cite{kn:brezhik00}.  In those cases the analogous formulae are
asymptotic approximations in the large-matrix limit.   The
expressions we obtain here are exact. Several stages of our work
were inspired by \cite{kn:basfor94} and \cite{kn:brezhik00}.  We
note that Fyodorov and Strahov
\cite{kn:fyostr03,kn:fyostr03a,kn:fyostr03b} have recently
extended the results of \cite{kn:andsim95} and \cite{kn:brezhik00}
for products as well as for ratios of shifted characteristic
polynomials of Hermitian matrices.

This paper is divided into three main sections, one devoted to
each of the three compact groups: $U(N)$, $O(2N)$ and $USp(2N)$.
In each we briefly present a related conjecture for the
autocorrelation functions for  families of $L$-functions having
the same unitary, orthogonal or symplectic symmetry. For more
details on the number theoretical side, see \cite{kn:cfkrs}.


\section{Unitary group: $U(N)$}

As mentioned in the introduction, we will calculate the
autocorrelation function
\begin{equation}
\label{eq1.1}
\int_{U(N)}\Lambda_M(s_1^{-1})\cdots \Lambda_M(s_m^{-1})
\Lambda_{M^{\dagger}}(s_{m+1})\cdots \Lambda_{M^{\dagger}}(s_{n})
dM,
\end{equation}
where $dM$ denotes Haar measure.  The
characteristic polynomial, which in this case we will
define as
\begin{equation}
\label{eq1.2}
\Lambda_M(s) =\det(I-Ms)= \prod_{n=1}^N (1-e^{i\theta_n}s),
\end{equation}
where
$e^{i\theta_n}$ are the eigenvalues of $M$, obeys the functional
equation
\begin{equation}
\label{eq1.3}
\Lambda_M(s)=(-1)^N \det M s^{N} \Lambda_{M^{\dagger}}(1/s),
\end{equation}
where
$M M^{\dagger}=I$.

We will actually examine
\begin{eqnarray}
\label{eq:avg}
I_{m,n}(U(N),w)&\equiv &I_{m,n}(U(N);w_1,\ldots,w_m;w_{m+1},\ldots,w_n)\nonumber \\
&=&\prod_{k=m+1}^nw_k^N\;\;\;\int_{U(N)}\prod_{i=m+1}^n \Lambda_M(w_i^{-1})\;\;\;
\prod_{j=1}^m\Lambda_{M^{\dagger}}(w_j)dM,
\end{eqnarray}
\noindent for which it transpires
that the result is simply interpreted through the work of
Nonnenmacher and Zirnbauer \cite{kn:nonzir02} as a character of
the group $U(n)$.  This is related to the correlation function (\ref{eq1.1})
via
\begin{eqnarray}
\label{eq:CI}
&&\int_{U(N)}\Lambda_M(s_1^{-1})\cdots \Lambda_M(s_m^{-1})\Lambda_{M^{\dagger}}(s_{m+1})
 \cdots \Lambda_{M^{\dagger}}(s_n) dM
\nonumber \\
&&\qquad\qquad\qquad\qquad=\left( \prod_{i=1}^{m} s_i^{-N}\right)\;
I_{n-m,n}(U(N);s_{m+1},\ldots,s_n;s_{1},
\ldots,s_{m}).
\end{eqnarray}

Our initial approach in this case (up to (\ref{eq:mn})) is
identical to \cite{kn:basfor94}.  We present this part of the
calculation in full, because it will be generalized in the
subsequent sections to the cases of $O(2N)$ and $USp(2N)$.

Using the expression for Haar measure in terms of the eigenvalues of $M$ \cite{kn:weyl},
\begin{eqnarray}
\label{eq:I}
I_{m,n}(U(N),w)&=& \frac{1}{N!(2\pi)^N} \int_0^{2\pi}\cdots \int_0^{2\pi}
 \left[\prod_{p=1}^N\left[
\left(\prod_{r=1}^m(1-e^{-i\theta_p}w_r)\right) \left(
\prod_{j=m+1}^n(w_j-e^{i\theta_p}) \right) \right]
\right]\nonumber \\
&&\qquad \times \prod_{1\leq \ell <q \leq
N}|e^{i\theta_q}-e^{i\theta_{\ell}}|^2d\theta_1\cdots d\theta_N.
\end{eqnarray}
Let $\Delta$ denote the Vandermonde determinant
\begin{equation}
\label{eq:vandermonde}
\Delta(x_1,\ldots,x_n)\equiv\prod_{1\leq j
<k \leq n} (x_k-x_j) = \det [x_j^{k-1}]_{1\leq j,k \leq
  n}.
\end{equation}
\noindent The object is to create in the integrand in (\ref{eq:I})
a Vandermonde determinant in the variables
$w_1,\ldots,w_n,e^{i\theta_1},\ldots,e^{i\theta_N}$. To this end
we introduce an extra factor $\prod_{1\leq \ell<m \leq k} (w_m-
w_{\ell})$, and, making use of the symmetry of the rest of the
integrand, replace $ \prod_{1\leq \ell <q \leq
N}|e^{i\theta_q}-e^{i\theta_{\ell}}|^2$ by
$(N!\prod_{j=1}^Ne^{-i(j-1)\theta_j}) \prod_{1\leq \ell <q \leq
N}(e^{i\theta_q}-e^{i\theta_{\ell}})$ in the integral.  This gives

\begin{eqnarray}
I_{m,n}(U(N),w)&= &\frac{(-1)^{(n-m)N}} {(2\pi)^N\prod_{1\leq \ell < q \leq n}
(w_{q}-w_{\ell}) } \int_0^{2\pi}\cdots \int_0^{2\pi}
 \left[ \prod_{p=1}^N e^{-im\theta_p}
\right] \nonumber \\
&&\qquad\times\left[ \prod_{p=1}^N\left(
\prod_{r=1}^m(e^{i\theta_p}-w_r)\right) \left(
\prod_{j=m+1}^n(e^{i\theta_p}-w_j)\right) \right]
\left(\prod_{1\leq \ell<q\leq
n}(w_q-w_{\ell})\right)\nonumber \\
&&\qquad \times \left(\prod_{1\leq \ell <q \leq N}
(e^{i\theta_q}-e^{i\theta_{\ell}})\right) \left( \prod_{j=1}^N
e^{-i(j-1)\theta_j} \right)d\theta_1\cdots d\theta_N \nonumber \\
&&=\frac{(-1)^{(n-m)N}} { (2\pi)^N\prod_{1\leq \ell < q \leq n}
(w_{q}-w_{\ell})} \int_0^{2\pi}\cdots \int_0^{2\pi}
 \left[ \prod_{j=1}^N
e^{-i(m+j-1)\theta_j}
\right]\nonumber \\
&&\qquad \times \left|\begin{array}{ccccc} 1& w_1 & w_1^2 & \cdots
& w_1^{N+n-1} \\ \vdots & \vdots & \vdots & \ddots & \vdots
\\ 1 & w_n & w_n^2 & \cdots & w_n^{N+n-1} \\ 1 & e^{i\theta_1} &
e^{2i\theta_1} & \cdots & e^{i(N+n-1)\theta_1} \\ \vdots & \vdots
& \vdots & \ddots & \vdots \\ 1 & e^{i\theta_N} & e^{2i\theta_N} &
\cdots & e^{i(N+n-1)\theta_N} \end{array} \right|d\theta_1\cdots d\theta_N.
\end{eqnarray}

If the factor $e^{-i(m+j-1)\theta_j}$ and the integration
over $\theta_j$ are pulled into the row of the determinant which
contains only $\theta_j$, then the integration in the final $N$
rows of the determinant results in zeros throughout these rows,
with the exception of a diagonal line of ones running from column
$m+1$ in row $n+1$ to column $m+N$ in row $n+N$.  Thus we are left
with the representation of $I$ as a determinant:
\begin{eqnarray}
\label{eq:mn}
&&I_{m,n}(U(N),w)\\
&&\quad=\frac{1}{\prod_{1\leq \ell <q\leq n}(w_q-w_{\ell})} \left|
\begin{array}{ccccccccc} 1 & w_1 & w_1^2 & \cdots & w_1^{m-1} &
w_1^{N+m} & w_1^{N+m+1} & \cdots & w_1^{N+n-1} \\ \vdots & \vdots
& \vdots & \ddots & \vdots & \vdots & \vdots & \ddots & \vdots \\
1 & w_n & w_n^2 & \cdots & w_n^{m-1} & w_n^{N+m} & w_n^{N+m+1} &
\cdots & w_n^{N+n-1}
\end{array}\right|.\nonumber
\end{eqnarray}
This result first appears in the work of Basor and Forrester
\cite{kn:basfor94}.

The notation can be simplified by recalling that the general form
of a Schur polynomial associated with the partition $\mu =
(\mu_1,\mu_2,\ldots,\mu_n)$ (where the $\mu_j$ are integers and
$\mu_1\geq \mu_2\geq \cdots \geq \mu_n\geq 0$) is
\begin{equation}
\label{eq:schur} S_{\mu}(x_1,\ldots,x_n) = \frac{\left|
\begin{array}{ccccc} x_1^{\mu_1+n-1} & x_1^{\mu_2+n-2} &
x_1^{\mu_3+n-3}& \cdots & x_1^{\mu_n} \\ x_2^{\mu_1+n-1} &
x_2^{\mu_2+n-2}& x_2^{\mu_3+n-3} &\cdots & x_2^{\mu_n} \\ \vdots &
\vdots & \vdots & \ddots & \vdots \\x_n^{\mu_1+n-1} &
x_n^{\mu_2+n-2}& x_n^{\mu_3+n-3} &\cdots & x_n^{\mu_n} \end{array}
\right| } { \left| \begin{array}{ccccc} x_1^{n-1} & x_1^{n-2} &
x_1^{n-3} & \cdots & 1 \\x_2^{n-1} & x_2^{n-2} & x_2^{n-3} &
\cdots & 1 \\ \vdots & \vdots & \vdots & \ddots & \vdots \\
x_n^{n-1} & x_n^{n-2} & x_n^{n-3} & \cdots & 1 \end{array}
\right|}.
\end{equation}
Thus,
\begin{equation}
I_{m,n}(U(N),w)=S_{\lambda^{(n-m)}} (w_1,\ldots,w_n),
\end{equation}
where $\lambda^{(n-m)}=(N,N,\ldots,N)$, with $(n-m)$ $N$'s.  This
is, as predicted from the approach of Zirnbauer and Nonnenmacher
\cite{kn:nonzir02} using Lie theory and dual pairs, a character of
an irreducible representation of the group $U(n)$ when
$w_1,\ldots,w_n$ lie on the unit circle.

We concentrate now on the determinant
\begin{eqnarray}
&&D_{N,m,n}(w_1, \ldots, w_{n})\nonumber \\
&&\qquad\equiv\left| \begin{array} {cccccccc} 1 & w_1 & w_1^2 & \cdots &
w_1^{m-1} & w_1^{N+m} & \cdots &
w_1^{N+n-1} \\
\vdots & \vdots& \vdots & \ddots& \vdots & \vdots& \ddots & \vdots \\
1& w_{n} & w_{n}^2 & \cdots & w_{n}^{m-1} & w_{n}^{N+m} & \cdots &
w_{n} ^{N+n-1} \end{array} \right| \nonumber \\
&& \qquad=\sum_{\sigma\in S_n} {\rm sgn}(\sigma) w_{\sigma(1)}^{0}
w_{\sigma(2)}^{1} w_{\sigma(3)}^{2}  \cdots w_{\sigma(m)}^{m-1}
w_{\sigma(m+1)}^{N+m} w_{\sigma(m+2)}^{N+m+1} \cdots
w_{\sigma(n)}^{N+n-1}, \label{eq:permut}
\end{eqnarray}
where the sum is over $S_n$, all permutations of
$\{1,2,\ldots, n\}$.  We break up the sum over all permutations into subsets.  Let
$\Xi_m$ be the set of the $\binom{n}{m}$ permutations
$\sigma\in S_{n}$ such that $\sigma(1)<\sigma(2)<\cdots<\sigma(m)$ and
$\sigma(m+1)<\cdots<\sigma(n)$.
\begin{eqnarray}
&&D_{N,m,n}(w_1, \ldots, w_{n})\nonumber \\
&&\qquad= \sum_{\sigma\in\; \Xi_m} {\rm sgn}(\sigma) \left(
\sum_{\rho}
    {\rm sgn}(\rho) w_{\rho(1)}^0 w_{\rho(2)}^1 \cdots
    w_{\rho(m)}^{m-1} \right) \left( \sum_{\delta}
    {\rm sgn}(\delta) w_{\delta(1)}^{0}w_{\delta(2)}^1 \cdots
    w_{\delta(n-m)}^{n-m-1} \right)\nonumber \\
&&\qquad \qquad\qquad\times (w_{\sigma(m+1)} w_{\sigma(m+2)} \cdots
    w_{\sigma(n)})^{N+m} ,
\end{eqnarray}
where $\rho$ is a permutation taking
    $\sigma(1),\sigma(2),\ldots,\sigma(m)$ to $\rho(1),\rho(2),\ldots,\rho(m)$ and
    $\delta$ is a permutation taking
    $\sigma(m+1),\sigma(m+2),\ldots,\sigma(n)$ to
    $\delta(1),\delta(2),\ldots,\delta(n-m)$.

Finally, using the definition of the Vandermonde determinant from
(\ref{eq:vandermonde}),
\begin{eqnarray}
\label{eq:finalD}
D_{N,m,n}(w_1, \ldots, w_{n})&=&\sum_{\sigma \in \;\Xi_m}
{\rm sgn}(\sigma) \left
  [ \prod_{1\leq \ell<j\leq m} (w_{\sigma(j)}-w_{\sigma( \ell)}) \right] \left
  [ \prod_{m+1\leq p<q\leq n}(w_{\sigma(q)}-w_{\sigma(p)})\right]\nonumber \\
&& \qquad\qquad\qquad\times (w_{\sigma(m+1)} w_{\sigma(m+2)} \cdots
    w_{\sigma(n)})^{N+m}.
\end{eqnarray}
So,
\begin{eqnarray}
&&I_{m,n}(U(N),w) = \sum_{\sigma \in \;\Xi_m} \frac {
(w_{\sigma(m+1)}w_{\sigma(m+2)} \cdots w_{\sigma(n)})^{N+m}}
{\prod_{{1 \leq \ell \leq m}\atop{m+1 \leq q \leq
        n}} (w_{\sigma(q)}-w_{\sigma(\ell)})  }.
\label{eq:result}
\end{eqnarray}
In (\ref{eq:finalD}) each factor $(w_i-w_j)$ is ordered such that
$i>j$.  In the denominator of (\ref{eq:result}) we wish the
ordering to be such that the first $w$ in each pair is chosen from
$w_{\sigma(m+1)}, \ldots, w_{\sigma(n)}$.  The sign required to accomplish
this reordering cancels exactly with ${\rm sgn} (\sigma)$ in the
numerator of (\ref{eq:finalD}).  Thus we obtain an expression for $I$
as a combinatorial sum:
\begin{eqnarray}
\label{eq:autocorr}
 I_{m,n}(U(N),w)&=&\sum_{\sigma \in \;\Xi_m}
 \frac { (w_{\sigma(m+1)}w_{\sigma(m+2)} \cdots
w_{\sigma(n)})^{N}}
{\prod_{{1 \leq \ell \leq m }\atop{ m+1 \leq q \leq
        n}} (1-w_{\sigma(\ell)}w_{\sigma(q)}^{-1})  } .
\end{eqnarray}

We now use \cite{kn:cfkrs}

\begin{lemma}
\label{lemma:unitary}

If
\begin{equation}
G(a_1,\ldots,a_m;b_1,\ldots,b_{n-m})=F(a_1,\ldots,a_m;b_1,\ldots,b_{n-m})
\prod_{i=1}^m \prod_{j=1}^{n-m} f(a_i-b_j),\nonumber
\end{equation}
\noindent where $F$ is regular near $(0,\ldots,0)$ and $f(x) =
\frac{1}{x} +c_0 +c_1 x + \cdots$, then
\begin{eqnarray}
&&\sum_{\sigma\in \Xi_m}
G(u_{\sigma(1)},\ldots,u_{\sigma(m)};u_{\sigma(m+1)},\ldots,u_{\sigma(n)} )\nonumber \\
&&\qquad\qquad= \frac{(-1)^{n(n-1)/2}}{(2\pi i)^{n} m! (n-m)!} \oint\cdots \oint
  G(z_1,\ldots,z_m;z_{m+1},\ldots, z_{n})\nonumber \\
&& \qquad\qquad\qquad\qquad\times\frac
{\Delta(z_1,\ldots,z_m,z_{m+1},\ldots,z_{n})^2} {\prod_{i=1}^n
\prod_{j=1}^{n}(z_i-u_j)} dz_1 \cdots dz_{n}, \nonumber
\end{eqnarray}
where $\Xi_m$ is the set of the $\binom{n}{m}$ permutations
$\sigma\in S_{n}$ such that $\sigma(1)<\sigma(2)<\cdots<\sigma(m)$ and
$\sigma(m+1)<\cdots<\sigma(n)$ and the contour integrals enclose the
variables $u_j$,
\end{lemma}

\noindent which allows us to write the sum (\ref{eq:autocorr})
as a contour integral:
\begin{eqnarray}
&&I_{m,n}(U(N);e^{-\alpha_1},e^{-\alpha_2},\ldots,e^{-\alpha_m};
e^{-\alpha_{m+1}}, \ldots ,e^{-\alpha_{n}})\nonumber \\
&&\qquad\qquad= \frac{(-1)^{n(n-1)/2}}{(2\pi i)^{n}m!(n-m)!} \oint\cdots \oint e^{-N
(z_{m+1}+z_{m+2}+\cdots +z_{n})} \prod_{{1\leq \ell\leq m}\atop{m+1\leq q\leq n}
}(1-e^{z_{q}-z_{\ell}})^{-1} \nonumber \\
&&\qquad\qquad\qquad\qquad\times \label{eq:RMTcontour} \frac{
\Delta(z_1,\ldots,z_m,z_{m+1},\ldots,z_{n})^2} { \prod_{i=1}^n
\prod_{j=1}^{n} (z_i-\alpha_j)} dz_1\cdots dz_{n}.
\end{eqnarray}
Br\'ezin and Hikami arrive at an integral of a very similar
form for the autocorrelation functions of
characteristic polynomials of random Hermitean matrices in the limit
of large matrix size $N$ \cite{kn:brezhik00}.
Note that in our case the result is an identity for any $N$.


\subsection{Comparison with the Riemann Zeta Function}
\label{sect:compRZF}

The main motivation for the calculations presented above
is to understand the autocorrelation function and moments of the
Riemann zeta function.  The Riemann zeta function is defined
for ${\rm Re} s >1$ by
 $\zeta(s)=\sum_{n=1}^{\infty} n^{-s}$ and has a
 continuation to a meromorphic function on the complex plane with a single,
 simple pole at $s=1$.  As described in detail in \cite{kn:cfkrs}, for
the autocorrelation functions of $\zeta(s)$ we have the following:
\begin{conjecture}
\begin{eqnarray*}
&&\int_0^T\zeta(\tfrac{1}{2}+\alpha_1+it)\cdots \zeta(\tfrac{1}{2}+\alpha_k+it)
\zeta(\tfrac{1}{2}-\alpha_{k+1}-it) \cdots \zeta(\tfrac{1}{2}
-\alpha_{2k}-it) dt \nonumber\\
&&\qquad \qquad= \int_0^{T}
W_k(t;\alpha_1,\ldots,\alpha_k;\alpha_{k+1},\ldots,\alpha_{2k})
(1+O(t^{-\tfrac{1}{2}+\epsilon})) dt,
\end{eqnarray*}
\noindent where
\begin{eqnarray}
\label{eq:W}
&&W_k(t;\alpha_1,\ldots,\alpha_k;\alpha_{k+1},\ldots,\alpha_{2k})
=e^{\frac{1}{2}\log \frac{t}{2\pi}(-\alpha_1-\alpha_2-\cdots -\alpha_k+\alpha_{k+1}
+\cdots +\alpha_{2k})}\nonumber \\
&& \qquad\qquad\qquad\qquad\times\sum_{\sigma\in \Xi} e^{\frac{1}{2}\log
\frac{t}{2\pi}(\alpha_{\sigma(1)}+\alpha_{\sigma(2)}+\cdots +
\alpha_{\sigma(k)}-\alpha_{\sigma(k+1)}-\cdots
-\alpha_{\sigma(2k)})} \nonumber\\
&&\qquad\qquad\qquad\qquad \times A_k(\alpha_{\sigma(1)},\ldots,\alpha_{\sigma(2k)})
\prod_{{1\leq \ell\leq k}\atop{k+1\leq m\leq 2k}}\zeta(1+\alpha_{\sigma(\ell)}-
\alpha_{\sigma(m)}),
\end{eqnarray}
\noindent and $\Xi$ is the set of the $\binom{2k}{k}$ permutations
$\sigma\in S_{2k}$ such that $\sigma(1)<\sigma(2)<\cdots<\sigma(k)$ and
$\sigma(k+1)<\cdots<\sigma(2k)$. Here  $A_k(u_1,\ldots,u_{2k})\equiv A_k(u)$
is an Euler
 product containing arithmetic information:
\begin{equation*}
A_k(u) =\prod_p \prod_{i=1}^k\prod_{j=1}^k
\left(1-\frac{1}{p^{1+u_i-u_{j+k}}}\right) \int_0^1 \prod_{j=1}^k
\left(1-\frac{e(\theta)}{p^{1/2+u_j}}\right)^{-1}
\left(1-\frac{e(-\theta)}{p^{1/2-u_{j+k}}}\right)^{-1}~d\theta.
\end{equation*}
\end{conjecture}

Note that by Lemma \ref{lemma:unitary} we can also write
\begin{eqnarray}
\label{eq:Wint}
&&W_k(t;\alpha_1,\ldots,\alpha_k;\alpha_{k+1},\ldots,\alpha_{2k})\nonumber \\
&&\qquad\qquad=e^{\frac{1}{2}\log \frac{t}{2\pi}(-\alpha_1-\alpha_2-\cdots -\alpha_k+\alpha_{k+1}
+\cdots +\alpha_{2k})}\frac{(-1)^k}{k!^2}\frac{1}{(2\pi i)^{2k}}\nonumber \\
&&\qquad\qquad\qquad\qquad \times\oint\cdots \oint
\frac{A_k(z_1,\dots,z_{2k})
\prod_{i=1}^k\prod_{j=1}^k\zeta(1+z_i-z_{j+k})\Delta(z_1,\dots,z_{2k})^2}
{ \prod_{i=1}^{2k}\prod_{j=1}^{2k} (z_i-\alpha_j)}\nonumber \\
&&\qquad\qquad\qquad\qquad \times\; e^{\frac{1}{2}\log
\frac{t}{2\pi}\sum_{j=1}^{k}z_j-z_{j+k}}~dz_1\dots dz_{2k} .
\end{eqnarray}

The Riemann zeta function satisfies a functional equation

\begin{equation}
\zeta(s)=\pi^{s-\tfrac{1}{2}}
\frac{\Gamma(\tfrac{1}{2}-\tfrac{1}{2}s)}{\Gamma(\tfrac{1}{2}
s)}\zeta(1-s).
\end{equation}

\noindent  The Riemann Hypothesis is that the complex zeros of $\zeta(s)$ lie on
the line ${\rm Re}s=1/2$.  The characteristic polynomial, on the
other hand, obeys the functional equation (\ref{eq1.3}) and its
zeros lie on the unit circle, so in analogy with the autocorrelation
functions of
$\zeta(s)$, we let $s_j=\exp(\alpha_j)$ in (\ref{eq:CI}).
Now when $\alpha_i$ is purely imaginary, $e^{-\alpha_i}$ sits on
the unit circle, in analogy with $1/2+it+\alpha_i$ lying on the
critical line when $\alpha_i$ is purely imaginary in the Riemann
zeta case.  We compare (\ref{eq:W}) with
\begin{eqnarray}
\label{eq:RMT}
 &&\int_{U(N)}\Lambda_M(e^{-\alpha_1})
\cdots \Lambda_M(e^{-\alpha_k})\Lambda_{M^{\dagger}}(e^{\alpha_{k+1}})\cdots
\Lambda_{M^{\dagger}}(e^{\alpha_{2k}})dM \nonumber \\
&&\qquad=  e^{\frac{N}{2}(-\alpha_1-\alpha_2-\cdots -\alpha_k+
\alpha_{k+1}+\cdots \alpha_{2k})}\\
&&\qquad\qquad \times \left(\sum_{\sigma \in \;\Xi}
e^{\frac{N}{2}(\alpha_{\sigma(1)}+\alpha_{\sigma(2)}+\cdots
+\alpha_{\sigma(k)}- \alpha_{\sigma(k+1)}
-\cdots -\alpha_{\sigma(2k)})}
\prod_{{1 \leq \ell \leq k }\atop{ k+1 \leq m \leq
        2k}} (1-e^{\alpha_{\sigma(m)}-\alpha_{\sigma(\ell)}})
        ^{-1}\right),\nonumber
\end{eqnarray}
which follows from (\ref{eq:CI}).  These two formulae clearly have a similar structure if we
 equate the density of the
Riemann zeros and the density of the eigenvalues of $M$ on the
unit circle to obtain the relation $N=\log \frac{t}{2\pi}$.  The
random matrix expression is, not surprisingly, missing the
arithmetical factor $A(\alpha_1,\ldots,\alpha_{2k})$; also, the
function which provides the simple poles in each term of the sum
is $\zeta(1+z)$ in the Riemann zeta case and $(1-e^{-z})^{-1}$ in
the random matrix case.


\section{Unitary symplectic group: $USp(2N)$}
\label{sect:Sp}

Now we turn to the group of symplectic unitary matrices,
$USp(2N)$. These are $2N\times 2N$ matrices, $M$, with
$MM^{\dagger}=1$ and $M^{t}JM=J$, where $J=\left(
  \begin{array} {cc}0&I_{N} \\-I_{N} &0 \end{array}\right)$ and $I_N$
is the $N\times N$ identity matrix.  For these matrices, the
eigenvalues lie on the unit circle and come in complex conjugate
pairs $e^{i\theta_1},e^{-i\theta_1},e^{i\theta_2},e^{-i\theta_2},
\ldots$ $e^{i\theta_N}, e^{-i\theta_N}$.  Thus we let the characteristic
polynomial related to such a
matrix take the form
\begin{eqnarray}
\Lambda_M(s)&=&\det(I-Ms)=\prod_{n=1}^N(1-e^{i\theta_n}s)(1-e^{-i\theta_n}s).
\end{eqnarray}
The weighting in the average over $USp(2N)$ of the matrix with
eigenphases $\pm \theta_1,\ldots,$ $\pm\theta_N$ is derived from
Haar measure on the group, and can be manipulated into the form
\begin{eqnarray}
&&N_{Sp} \frac{(-1)^{N(N-1)/2}
}{4^{N^2}}\Delta(e^{i\theta_1},\ldots,e^{i\theta_N},
e^{-i\theta_1},\ldots,e^{-i\theta_N})
\prod_{k=1}^N(e^{i\theta_k}-e^{-i\theta_k}),
\end{eqnarray}
\noindent where $N_{Sp}=\frac{2^{2N^2-2N}}{\pi^N N!}$.

We define the autocorrelation function in this case to be
\begin{eqnarray}
&&I(USp(2N),w_1,\ldots,w_k)\equiv \int_{USp(2N)}\Lambda_M(w_1)\cdots
\Lambda_M(w_k) dM
\nonumber \\
&&\qquad=\frac{N_{Sp}(-1)^{N(N-1)/2}}{4^{N^2}}
\int_{0}^{2\pi}\cdots\int_{0}^{2\pi}
\Delta(e^{i\theta_1},\ldots,e^{i\theta_N},e^{-i\theta_1},\ldots,
e^{-i\theta_N})\nonumber
\\
&&\qquad \qquad\times
\prod_{m=1}^k\prod_{n=1}^{N}(e^{i\theta_n}-w_m)
(e^{-i\theta_n}-w_m)\prod_{j=1}^N
(e^{i\theta_j}-e^{-i\theta_j}) d\theta_1 \cdots d\theta_N\nonumber \\
&&\qquad=\frac{N_{Sp}}{4^{N^2}}\frac{(-1)^{N(N-1)/2}}{\prod_{1\leq
i<j\leq k} (w_j-w_i)} \int_{0}^{2\pi} \cdots \int_{0}^{2\pi}
  \nonumber \\
&&\qquad\qquad\times
\Delta(w_1,w_2,\ldots,w_k,e^{i\theta_1},\ldots ,e^{i\theta_N},
e^{-i\theta_1},\ldots,e^{-i\theta_N})
\prod_{j=1}^N(e^{i\theta_j}-e^{-i\theta_j})d\theta_1\cdots d\theta_N \nonumber \\
&&\qquad=\frac{N_{Sp}}{4^{N^2}} \frac{(-1)^{N(N-1)/2}}{\prod_{1\leq
i<j\leq k} (w_j-w_i)}\int_{0}^{2\pi}\cdots \int_{0}^{2\pi}  \nonumber \\
&&\qquad \qquad\times \sum_{\sigma\in S_{k+2N}}{\rm sgn}(\sigma) w_1^{\sigma(1)-1}
w_2^{\sigma(2)-1} \cdots w_k^{\sigma(k) -1}
(e^{i\sigma(k+1)\theta_1} - e^{i(\sigma(k+1)-2)\theta_1 })
\nonumber \\
&& \qquad \qquad\times (e^{i\sigma(k+2)\theta_2}-e^{i(\sigma(k+2)-2)\theta_2}) \cdots
(e^{i\sigma(k+N)\theta_N} - e^{i(\sigma(k+N)-2)\theta_N}) \nonumber \\
&&\qquad\qquad \times e^{-i(\sigma(k+N+1)-1)\theta_1} \cdots
e^{-i(\sigma(k+2N)-1)\theta_N}d
\theta_1\cdots d \theta_N.\label{eq:sigmaSp}
\end{eqnarray}

As we are integrating each $\theta_j$ from 0 to $2\pi$, the term in
the sum belonging to a given permutation $\sigma$ is zero unless
for every $j$, $\sigma(k+j)=\sigma(k+N+j)-1$ or
$\sigma(k+j)=\sigma(k+N+j)+1$.  Upon integration this places the
condition $i_1< i_2< \cdots < i_k \in \{0,1,2,\ldots,2N+k-1\}$,
$i_j$ is even if $j$ is odd and $i_j$ is odd if $j$ is even, on
the resulting sum over $k\times k$ determinants:
\begin{eqnarray}
\label{eq:detsum}
&&I(USp(2N),w_1,\ldots,w_k)\nonumber \\
&&\qquad\qquad= \frac{1}{\prod_{1\leq i<j\leq k}(w_j-w_i)}
\sum_{{0\leq i_1 <i_2 <\cdots <i_k\leq 2N+k-1}\atop{ i_j\equiv j-1
mod 2}}\left|\begin{array}{cccc} w_1^{i_1}&w_1^{i_2}
&\cdots& w_1^{i_k} \\
w_2^{i_1}&w_2^{i_2}&\cdots &w_2^{i_k}\\ \vdots&\vdots&
\ddots&\vdots \\ w_k^{i_1}&w_k^{i_2}&\cdots&w_k^{i_k} \end{array}
\right|.
\end{eqnarray}
Note that this can also be written in terms of Schur functions
(see (\ref{eq:schur})),
\begin{eqnarray}
\label{eq:schur2} &&I(USp(2N),w_1,\ldots,w_k)=\sum_{\lambda {\rm
\; even}} S_{\lambda}(w_1,\ldots,w_k),
\end{eqnarray}
\noindent where the sum is over partitions
$\lambda=(\lambda_1,\ldots,\lambda_k)$ with all parts $\lambda_j$
even and $2N\geq \lambda_1 \geq \lambda_2 \geq \cdots \geq
\lambda_k \geq 0$.

Examination of examples when $k$ is small leads to the guess that in general,
\begin{eqnarray}
\label{eq:kth} && I(USp(2N),w_1,\ldots,w_k)\nonumber \\
&&\qquad\qquad= w_1^N\cdots w_k^N \left[ \sum_{\epsilon_j\in \{-1,1\}}
(\prod_{j=1}^k w_j^{\epsilon_jN})  \prod_{1\leq i\leq j \leq k}
(1- w_i^{-\epsilon_i}w_j^{-\epsilon_j})^{-1} \right] ,
\end{eqnarray}
and we will now prove this to be true.

Before embarking on the proof of (\ref{eq:kth}) we note that
 letting $w_j^N=e^{b_j}$ and taking $N$ large,
\begin{eqnarray}
&&I(USp(2N),e^{b_1/N},\ldots,e^{b_k/N} )\nonumber \\
&&\qquad \qquad\approx e^{b_1}\cdots e^{b_k} \left[
\sum_{\epsilon_j\in\{-1,1\}} \left( \prod_{j=1}^k
e^{\epsilon_jb_j}  \right)  \prod_{1 \leq i\leq j \leq k} \left(
\frac{\epsilon_i b_i}{N} + \frac{\epsilon_j b_j} {N} \right)^{-1}
\right] \nonumber \\
&&\qquad\qquad= N^{ \frac{k^2+k}{2}} e^{b_1} \cdots e^{b_k} \left[
\sum_{\epsilon_j\in\{-1,1\}} \left( \prod_{j=1}^k e^{\epsilon_j
b_j} \right) \prod_{1\leq i\leq j \leq k}
(\epsilon_ib_i+\epsilon_jb_j)^{-1}\right] .
\end{eqnarray}
The sum here has just the same structure as Br\'ezin and Hikami's
results for the large $N$ asymptotics of Hermitian ensembles
\cite{kn:brezhik00}, showing that
when distances are measured in terms of the mean level spacing of the
eigenvalues then, as expected, in the large $N$ limit averages over the
compact groups and the Hermitian ensembles are equivalent.

To prove (\ref{eq:kth}), we need two indentities. The first is
\begin{identity}
\label{iden:1}

\begin{eqnarray}
 &&\sum_{j=1}^n
\Delta(w_1,\ldots,w_n)\big|_{w_j=0} \prod_{m=1}^n
(1-w_jw_m)=(1-w_1^2\cdots w_n^2)\Delta(w_1,\cdots,w_n).\nonumber
\end{eqnarray}

\end{identity}
This is a special case, with $f(w)=\prod_{m=1}^n(1-w_mw)$, of the
following lemma:

\begin{lemma}
\label{lemma1}

Given a polynomial function of order $n$,
$f(w)=c_0+c_1w+\cdots+c_nw^n$, we have the relation
\begin{eqnarray}
\label{eq:lemma1} && \sum_{j=1}^n
\Delta(w_1,\ldots,w_n)\big|_{w_j=0} f(w_j) \nonumber\\
&&\qquad\qquad=  (c_0+(-1)^{n-1}c_nw_1\cdots w_n)\Delta(w_1,\ldots,w_n) .\nonumber
\end{eqnarray}
\end{lemma}

To prove Lemma \ref{lemma1} we notice first of all that we can
write the left side of the relation as a determinant.

\begin{equation}
\label{eq:fdet}
\sum_{j=1}^n \Delta(w_1,\ldots,w_n)\big|_{w_j=0} f(w_j)=\left|
\begin{array}{ccccc}f(w_1)&w_1&w_1^2&\cdots &w_1^{n-1} \\ f(w_2)&
w_2 & w_2^2&\cdots & w_2^{n-1} \\ \vdots& \vdots & \vdots & \ddots
& \vdots\\ f(w_n) & w_n& w_n^2 & \cdots & w_n^{n-1} \end{array}
\right|.
\end{equation}
However, since $f(w)$ is a polynomial of order $n$, in the first
column of the above determinant, all the terms in $f$ with
coefficients $c_1,\ldots,c_{n-1}$ can be cancelled by column
manipulations, leaving just
\begin{eqnarray}
\sum_{j=1}^n \Delta(w_1,\ldots,w_n)\big|_{w_j=0} f(w_j)&=&\left|
\begin{array}{cccc} c_0+c_nw_1^n & w_1 & \cdots & w_{1}^{n-1} \\
c_0+c_nw_2^n & w_2& \cdots & w_2^{n-1} \\ \vdots & \vdots & \ddots
& \vdots \\ c_0+c_nw_n^n & w_n & \cdots & w_n^{n-1} \end{array}
\right|\nonumber \\
&=& (c_0+(-1)^{n-1}c_nw_1\cdots w_n)\Delta(w_1,\ldots,w_n).
\end{eqnarray}

The second identity is

\begin{identity}
\label{iden:2}

\begin{eqnarray}
&& \sum_{C\cup D=[n],C\cap D=\emptyset} (-1)^{S(C,D)}
 \; \prod_{\alpha\in C}
w_{\alpha}^{n-1} \;\prod_{{i,j\in C}\atop{i<j}}(w_j-w_i)\;
\prod_{{i,j\in D}\atop{i<j}}(w_j-w_i)\;
 \prod_{{\alpha\in C }\atop{
\beta\in D }} (1-w_{\alpha}w_{\beta}) =0,\nonumber
\end{eqnarray}

\end{identity}
\noindent where the left hand side is a polynomial in the
variables $w_1,\ldots, w_n$ and the notation is $[n]=\{1,2,\ldots,n\}$,
$|C|$ is the number of elements in $C$,
$W(C,D)=\sum_{{m\in C, \;n\in D}\atop{m>n}} 1$ and
$S(C,D)=|C||D|+|C|(|C|+1)/2 +W(C,D)$.

We prove this by showing that when $r=n-1$, with the same notation
as above,
\begin{eqnarray}
\label{eq:iden2x}
&&F_n(w_1,\ldots,w_n;x;r)\\
&& \qquad=\sum_{C\cup D=[n],C\cap D=\emptyset} (-1)^{S(C,D)}
 \; \prod_{\alpha\in C} w_{\alpha}^{r}\;\;
\;\Delta(C) \Delta(D) \; \prod_{{ \alpha\in C}\atop{ \beta\in D}
}(x^2-w_{\alpha}w_{\beta}) \;\;\; x^{|D|^2+(r-n)|D| }\nonumber
\end{eqnarray}
\noindent is identically zero.  Here $\Delta(C)=\prod_{{i,j\in C}
\atop{i<j}}(w_j-w_i)$.  We proceed by showing that the
polynomial $F_n(w_1,\ldots,w_n;x;n-1)$, which is of order $n(n-1)$
in $x$, has at least $n(n-1)+1$ roots, implying that it is
identically zero. Since the left hand side of the equation in Identity
\ref{iden:2} is
merely the instance of $F_n(w_1,\ldots,w_n;x;n-1)$ when $x=1$,
this proves Identity \ref{iden:2}.

First of all we note that $F_n(w_1,\ldots,w_n;x;n-1)$ is zero when
$x$ is zero. Only the terms with $|D|=0,1$ contribute
in this case, due to the factor
$x^{|D|^2-|D| }$. Thus we are looking at
\begin{eqnarray}
(-1)^{n(n+1)/2}\Delta(w_1,\ldots,w_n) \prod_{i=1}^n w_i^{n-1}
+(-1)^{n(n+1)/2-1}\sum_{j=1}^n \Delta(w_1,\ldots,w_n)\big|_{w_j=0}
\prod_{i=1}^n w_i^{n-1},
\end{eqnarray}
\noindent which we can see is zero by a simple application of
Lemma \ref{lemma1} (with $f(w)=(-1)^{n(n+1)/2}$ $\times
\prod_{i=1}^nw_i^{n-1}$).

Next we prove that (\ref{eq:iden2x}) is zero for certain values of
the integer $r\leq n-1$ when $x^2=w_aw_b$, with $a\neq b =
1,2,\ldots,n$.  This yields $n(n-1)$ other zeros (assuming none of
the $w_j$ are zero) and proves that (\ref{eq:iden2x}) is
identically zero for $r=n-1$. We start with
$F_n(w_1,\ldots,w_n;\sqrt{w_aw_b};r)$.  We note immediately that
in the sum over $C$ and $D$, any term in which $a$ and $b$ do not
occur both in $C$ or both in $D$ is zero. Thus,
\begin{eqnarray}
\label{eq:I1}
 &&F_n(w_1,\ldots,w_n;\sqrt{w_aw_b};r) \\
 &&\qquad\qquad= \sum_{A\cup B=[n]_{a,b},\;A\cap B=\emptyset} (-1)^{S(A,B\cup\{a,b\})} \;
\prod_{\alpha\in A} w_{\alpha}^{r} \;\;\; \Delta(A)
\Delta(B\cup\{a,b\})\;\nonumber \\
&&\qquad\qquad\qquad\qquad \times \prod_{{\alpha \in A}\atop { \beta\in
B\cup\{a,b\}}} (w_aw_b-w_{\alpha}w_{\beta})\;\;\;(\sqrt{w_aw_b})^
{|B\cup\{a,b\}|^2+(r-n)|B\cup\{a,b\}| }\nonumber \\
&& \qquad\qquad \qquad\qquad+ \sum_{A\cup B=[n]_{a,b},\;A\cap B=\emptyset}
(-1)^{S(A\cup\{a,b\},B)} \; \prod_{\alpha\in A\cup\{a,b\}}
w_{\alpha}^{r}\;\;
\;\Delta(A\cup\{a,b\}) \Delta(B)\;\nonumber \\
&&\qquad\qquad\qquad\qquad \times \prod_{{ \alpha \in A\cup\{a,b\}} \atop{ \beta\in
B} }(w_aw_b-w_{\alpha}w_{\beta})\;\;\;(\sqrt{w_aw_b})^
{|B|^2+(r-n)|B| },\nonumber
\end{eqnarray}
\noindent where $[n]_{a,b}$ is the set of elements $\{1,2,\ldots,n\}$ with
$a$ and $b$ removed.
However, after some manipulations we can write both the sum in
(\ref{eq:I1}) containing $A$ and $B\cup\{a,b\}$ and the sum
containing $A\cup\{a,b\}$ and $B$ in terms of just $A$ and
$B$.  To this end, we note that
$|B\cup\{a,b\}|=|B|+2$, $S(A\cup \{a,b\},B)=(S(A,B)+s_b+1)\mod 2\;$ and $
S(A,B\cup\{a,b\}) = (S(A,B)+s_a)\mod 2$, where (assuming $a>b$) $s_A$ is
$\sum_{i\in A,\; a>i>b} 1$  and $s_B$ is $\sum_{i\in B,\;a>i>b} 1$.   Hence
\begin{eqnarray}
&&F_n(w_1,\ldots,w_n;\sqrt{w_aw_b};r) \nonumber \\
&& \qquad= \sum_{A\cup B=[n]_{a,b},\;A\cap B=\emptyset}(-1)^{S(A,B)+s_A+s_B}
 \; \prod_{\alpha\in A} w_{\alpha}^{r} \;\;\;
\Delta(A) \Delta(B)\nonumber \\
&&\qquad\qquad \times \left( \prod_{\beta\in B}
(w_a-w_{\beta})(w_b-w_{\beta}) \right) (w_a-w_b)\;
\prod_{{ \alpha \in A}\atop
{ \beta\in B}} (w_aw_b-w_{\alpha}w_{\beta})\;\nonumber \\
&&\qquad\qquad \times \prod_{\alpha\in A}
 (w_b-w_{\alpha}) (w_a-w_{\alpha}) \;\;\;(\sqrt{w_aw_b})^
{|B|^2+(r-n)|B|
}\;(w_aw_b)^{|A|}(w_aw_b)^{2|B|
+(r-n)+2}\nonumber \\
&&\qquad\qquad +  \sum_{A\cup B=[n]_{a,b},\;A\cap B=\emptyset}(-1)^{S(A,B) +s_A+s_B+1}
 \;  \prod_{\alpha\in A} w_{\alpha}^{r}\;\;
\;\Delta(A) \Delta(B) \nonumber \\
&&\qquad \qquad\times \left( \prod_{\alpha\in A}
(w_a-w_{\alpha})(w_b-w_{\alpha}) \right) (w_a-w_b)\;
\prod_{{\alpha \in A} \atop{ \beta\in B}
} (w_aw_b-w_{\alpha}w_{\beta})\; \nonumber \\
&&\qquad \qquad\times\prod_{\beta\in B} (w_b-w_{\beta})(w_a-w_{\beta})
\;\;\;(\sqrt{w_aw_b})^ {|B|^2+(r-n)|B| }(w_a w_b)^r(w_aw_b)^{|B|}.
\end{eqnarray}
Since $|A|+|B|+2=n$, we see that the two sums
above cancel each other exactly, term by term.  Thus we have that
\begin{equation}
F_n(w_1,\ldots,w_n;\sqrt{w_aw_b};r) =0.
\end{equation}
If $n-r$ is odd, it immediately follows that
$F_n(w_1,\ldots,w_n;-\sqrt{w_aw_b};r)=0$ also, as
(\ref{eq:iden2x}) will be even in $x$.  This proves Identity
\ref{iden:2}.

Since the proof of $F_n(w_1,\ldots,w_n;\sqrt{w_aw_b};r) =0$
involved cancellation in (\ref{eq:iden2x}) only amongst terms in
which $|C|$ has the same parity, we can restrict the sum over $C$
to sets of even cardinality or sets of odd cardinality.  Note then
that if $r=n-2$, we can write a further identity (which will be of
use in Section \ref{sect:SO2N})

\begin{identity}
\label{iden:3}

\begin{equation}
\sum_{{C\cup D=[n],C\cap D=\emptyset}\atop {|C| {\rm \; even}}}
(-1)^{S(C,D)}\; \prod_{\alpha\in C} w_{\alpha}^{n-2}\;\;
\;\Delta(C) \Delta(D) \; \prod_{{ \alpha\in C}\atop{ \beta\in D}}
(x^2-w_{\alpha}w_{\beta}) \;\;\; x^{|D|^2-2|D|+1 }=0,\nonumber
\end{equation}

\end{identity}
\noindent because in the same manner as above, we would see that
the left side of the expression is zero when $x=+\sqrt{w_aw_b}$,
$a\neq b = 1,2,\ldots,n$.  Note that an extra factor of $x$ has
been included in each term to ensure that the expression on the
left of Identity \ref{iden:3} is a polynomial in $x$; that is,
there are no terms with negative exponents on $x$.  To deal with
$x=-\sqrt{w_aw_b}$, we note that
 if $n$ is odd, the expression is an even
polynomial in $x$, and if $n$ is even, then $x^{|D|^2+2|D|+1}$ is
always an odd power of $x$. Thus the expression is zero when
$x=\pm \sqrt{w_aw_b}$, $a\neq b=1,2,\ldots,n$ (this means we have
$n(n-1)$ zeros), and the polynomial in $x$ is of order
$n(n-1)-n+1$.  Thus if $n\geq 2$, it is everywhere zero and
Identity \ref{iden:3} is true.

We are now in a position to return to the proof of (\ref{eq:kth}).
We need to prove that this is identical to
(\ref{eq:detsum}).  We will now prove that
\begin{eqnarray}
\label{eq:theorem}
 &&\hspace{-0.5 in}\frac{1}{\Delta(w_1,\ldots,w_k)} \sum_{{0\leq i_1 <i_2
<\cdots <i_k\leq n}\atop{ i_j\equiv j-1 mod
2}}\left|\begin{array}{cccc} w_1^{i_1}&w_1^{i_2}
&\cdots& w_1^{i_k} \\
w_2^{i_1}&w_2^{i_2}&\cdots &w_2^{i_k}\\ \vdots&\vdots&
\ddots&\vdots \\ w_k^{i_1}&w_k^{i_2}&\cdots&w_k^{i_k} \end{array}
\right|\\
&&\hspace{-0.2 in}=w_1^{(n-k+1)/2}\cdots w_k^{(n-k+1)/2}
\sum_{\epsilon_j\in \{-1,1\}} \left(\prod_{m=1}^k w_m^{\epsilon_m
(n-k+1)/2}\right)\left( \prod_{1\leq m\leq q\leq k }
(1-w_m^{-\epsilon_m}w_q^{-\epsilon_q})^{-1}\right).\nonumber
\end{eqnarray}

As a first step it is convenient to add to the notation already
introduced to help simplify the
equations.  If $A$ and $B$ are sets of positive integers, then we
let $w_A=\prod_{m\in A}w_m$.  Further, we define
\begin{equation}
E(A)=\prod_{{m\leq n}\atop {m,n\in A}}(1-w_mw_n)
\end{equation}
\noindent as well as
\begin{equation}
E(A,B)=\prod_{{m\in A}\atop{n\in B}} (1-w_mw_n),
\end{equation}
\noindent and, as previously,
\begin{equation}
\Delta(A)=\prod_{{m<n}\atop{m,n\in A}} (w_n-w_m)
\end{equation}
\noindent and
\begin{equation}
D(A,B)=\prod_{{m\in A}\atop{n\in B}}(w_n-w_m).
\end{equation}

Armed with this notation, the right side of
(\ref{eq:theorem}) can be written, where $A$ is the set of indices
$j$ for which $\epsilon_j=+1$, as
\begin{eqnarray}
&&w_1^{(n-k+1)/2}\cdots w_k^{(n-k+1)/2} \sum_{A\cup B= [k],\;
A\cap B=\emptyset}w_A^{(n-k+1)/2}w_{B}^{-(n-k+1)/2} \nonumber \\
&& \qquad\qquad\qquad \qquad\times
 \left(\prod_{{m,q\in A}\atop{m\leq q}}
\frac{1}{1-\frac{1}{w_mw_q}} \right)\; \left(\prod_{{m,q\in
B}\atop {m\leq q}} \frac{1}{1-w_mw_q}\right) \; \left(\prod_{{m\in
A}\atop {q\in
B}} \frac{1}{1-w_q/w_m}\right)\nonumber \\
&&\qquad\qquad=\sum_{A\cup B= [k],\; A\cap
B=\emptyset}\frac{w_A^{n-k+1} }{E(A)E(B) D(A,B)}
\left(\prod_{{m,q\in A}\atop{m\leq q}} -w_mw_q\right) \;\left(
\prod_{{m\in A}\atop {q\in B}} -w_m\right).
\end{eqnarray}
A straightforward manipulation gives
\begin{eqnarray}
\prod_{{m,n\in A}\atop {m\leq n}} -w_mw_n &=& w_A^{|A
|+1}(-1)^{|A|(|A|+1)/2}, \nonumber \\
\prod_{{m\in A}\atop {n\in B}}-w_m &=&
w_A^{|B|}(-1)^{|A||B|}\;\;{\rm and} \nonumber \\
E(A\cup B)&=&E(A)E(B)E(A,B) \;\;{\rm and \;similarly}\nonumber \\
\Delta(A\cup B)&=&\Delta(A)\Delta(B)D(A,B)(-1)^{W(A,B)},
\end{eqnarray}
\noindent so we arrive at a re-expression of (\ref{eq:theorem}):
\begin{eqnarray}
\label{eq:theorem2}
 &&\sum_{{0\leq i_1 <i_2 <\cdots <i_k\leq
n}\atop{ i_j\equiv j-1 mod 2}}\left|\begin{array}{cccc}
w_1^{i_1}&w_1^{i_2}
&\cdots& w_1^{i_k} \\
w_2^{i_1}&w_2^{i_2}&\cdots &w_2^{i_k}\\ \vdots&\vdots&
\ddots&\vdots \\ w_k^{i_1}&w_k^{i_2}&\cdots&w_k^{i_k} \end{array}
\right|\nonumber \\
&&\qquad\qquad=\frac{1}{E([k])} \sum_{A\cup B=[k],\;A\cap B=\emptyset}
(-1)^{S(A,B)} w_A^{n+2} E(A,B)
\Delta(A)\Delta(B).
\end{eqnarray}

We prove this by induction on $k$.  We see that when
$k=1$
\begin{equation}
\sum_{{0\leq i_1\leq n}\atop{i_1 \;{\rm even}}} w_1^{i_1}=
\frac{1-w_1^{n+2}}{1-w_1^2},
\end{equation}
\noindent which clearly satisfies (\ref{eq:theorem2}).  We now
show that if (\ref{eq:theorem2}) holds with $k$ replaced by $k-1$,
then it holds for $k$ as well.  We start by expanding the
determinant in (\ref{eq:theorem2}) with respect to the last column
so that the left side becomes
\begin{eqnarray}
&&\sum_{{k-1\leq i_k \leq n}\atop {i_k\equiv k-1 \; {\rm mod}
\;2}} \sum_{j=1}^k w_j^{i_k} (-1)^{k-j} \sum_{{0\leq i_1 < i_2
<\cdots < i_{k-1} \leq i_k-1} \atop {i_r \equiv r-1 \;{\rm
mod}\;2}} \det\big( w_m^{i_q}\big)_{{1\leq m\leq k,\; m\neq
j}\atop{1 \leq q\leq k-1}}.
\end{eqnarray}
By the induction hypothesis, this is
\begin{eqnarray}
\label{eq:sets1}
 &&\hspace{-0.4 in}\sum_{{k-1\leq i_k \leq n}\atop {i_k\equiv
k-1\;{\rm mod} \;2}} \sum_{j=1}^k w_j^{i_k} (-1)^{k-j}
\frac{1}{E([k]_j)} \sum_{{A\cup B =[k]_j}\atop{A\cap B=\emptyset}}
(-1)^{S(A,B)} w_A^{i_k+1} E(A,B) \Delta(A) \Delta(B),
\end{eqnarray}
\noindent where $A_j=A-\{j\}$.

If we redefine $A$ to include $j$, and switch the order of the sum
over $j$ and the sum over the sets $A$ and $B$ in
(\ref{eq:sets1}), we obtain
\begin{eqnarray}
&&\hspace{-0.2 in}\sum_{{k-1\leq i_k \leq n}\atop {i_k\equiv
k-1\;{\rm mod} \;2}}\sum_{{A\cup B=[k]}\atop{A\cap B=\emptyset}}
\sum_{j\in A} \frac{(-1)^{k-j} w_j^{i_k} }{E([k]_j)}
(-1)^{S(A_j,B)} w_{A_j}^{i_k+1} E(A_j,B)\Delta(A_j)\Delta(B).
\end{eqnarray}
Applying the definition of $E$, it is straightforward to show that
for $A\cup B=[k]$, $A\cap B = \emptyset$ and $j\in A$, then
$E(A_j,B)/E([k]_j)=E(A,B)E(\{j\},A)/E([k])$.  This leads us to
\begin{eqnarray}
\label{eq:sets2}
 &&\sum_{{k-1\leq i_k \leq n}\atop {i_k\equiv
k-1\;{\rm mod} \;2}}\frac{1}{E([k])} \sum_{{A\cup B=[k]}\atop{A\cap
B=\emptyset}} w_A^{i_k}E(A,B)\Delta(B) \nonumber \\
&&\qquad\qquad \qquad \times\sum_{j\in A}  (-1)^{k-j}
(-1)^{S(A_j,B)} w_{A_j} E(\{j\},A)\Delta(A_j).
\end{eqnarray}
\noindent In this notation, a simple generalization of Identity
\ref{iden:1} can be written as
\begin{equation}
\sum_{j\in A} (-1)^{W(\{j\},A)} w_{A_j} E(\{j\},A) \Delta(A_j) =
\Delta(A)(1-w_A^2),
\end{equation}
\noindent and this combined with
$(-1)^{S(A_j,B)+k-j}=(-1)^{W(\{j\},A)} (-1)^{S(A,B)+1}$, where
$A\cup B=[k]$, $A\cap B=\emptyset$ and $j\in A$, gives us
\begin{eqnarray}
&&\hspace{-0.2 in} \sum_{{k-1\leq i_k \leq n}\atop {i_k\equiv
k-1\;{\rm mod} \;2}} \frac{1}{E([k])} \sum_{A\cup B=[k],A\cap
B=\emptyset} (-1)^{S(A,B)+1} \Delta(B) E(A,B) w_{A}^{i_k}
\Delta(A)(1-w_A^2).
\end{eqnarray}
\noindent Summing over $i_k$, this yields
\begin{eqnarray}
&& \frac{1}{E([k])} \sum_{A\cup B=[k],A\cap B=\emptyset} (-1)^{S(A,B)}
\Delta(A) \Delta(B) E(A,B)
\big[w_A^{n+2}-w_A^{k-1}\big].
\end{eqnarray}

Applying Identity \ref{iden:2}, we see that the terms resulting
from $w_A^{k-1}$ in the square brackets above all cancel out,
leaving us with
\begin{equation}
\frac{1}{E([k])} \sum_{A\cup B=[k],A\cap B=\emptyset}
 \Delta(A)\Delta(B)E(A,B)w_A^{n+2} (-1)^{S(A,B)},
\end{equation}
\noindent which proves (\ref{eq:theorem2}) and so proves (\ref{eq:kth}).

We also have the following lemma
\cite{kn:cfkrs}

\begin{lemma}
\label{lemmaint}

If $F$ is a symmetric function of $k$ variables, regular near
$(0,\ldots,0)$, and $f(x)$ has a simple pole of residue 1 at $x=0$
and is otherwise analytic in a neighbourhood of $x=0$,  and either
\begin{equation}
G(a_1,\ldots,a_k)=F(a_1,\ldots,a_k) \prod_{1\leq i \leq j\leq k}
f(a_i+a_j)
\end{equation}
\noindent or
\begin{equation}
G(a_1,\ldots,a_k)=F(a_1,\ldots,a_k) \prod_{1\leq i < j\leq k}
f(a_i+a_j),
\end{equation}
\noindent then when $\pm\alpha_i\pm\alpha_j$ are contained in the
region of analyticity of $f(x)$
\begin{eqnarray}
&&\hspace{-0.2 in}\sum_{\epsilon_j \in \{-1,1\}} G(\epsilon_1 \alpha_1,\ldots,
\epsilon_k\alpha_k) \\
&&\hspace{-0.2 in}\qquad\qquad=\frac{(-1)^{k(k-1)/2} } {(2\pi i)^k}
\frac{2^k}{k!} \oint\cdots \oint G(z_1,\ldots,z_k)
\frac{\Delta(z_1^2,\ldots,z_k^2)^2 \prod_{j=1}^k z_j }
{\prod_{i=1}^k\prod_{j=1}^k (z_i-\alpha_j)(z_i+\alpha_j)}
dz_1\cdots dz_k \nonumber
\end{eqnarray}
\noindent and
\begin{eqnarray}
&&\hspace{-0.2 in}\sum_{\epsilon_j \in \{-1,1\}} (\prod_{j=1}^k
\epsilon_j)G(\epsilon_1 \alpha_1,\ldots,
\epsilon_k\alpha_k)  \\
&&\hspace{-0.2 in}\qquad\qquad=\frac{(-1)^{k(k-1)/2} } {(2\pi i)^k}
\frac{2^k}{k!} \oint \cdots \oint G(z_1,\ldots,z_k)
\frac{\Delta(z_1^2,\ldots,z_k^2)^2 \prod_{j=1}^k \alpha_j }
{\prod_{i=1}^k\prod_{j=1}^k (z_i-\alpha_j)(z_i+\alpha_j)}
dz_1\cdots dz_k \nonumber,
\end{eqnarray}
\noindent where the contour of integration encircles the $\pm
\alpha$'s.

\end{lemma}

With the help of Lemma \ref{lemmaint} we can write
\begin{eqnarray}
\label{eq:SpBH}
&&I(USp(2N),e^{-\alpha_1},\ldots,e^{-\alpha_k})\nonumber \\
&&\qquad \qquad=\frac{(-1)^{k(k-1)/2} 2^k} {(2\pi i)^k k!}
e^{-N\sum_{j=1}^k \alpha_j} \oint \cdots \oint \prod_{1\leq \ell
\leq m \leq k} (1-e^{-z_m-z_\ell})^{-1} \nonumber \\
&& \qquad\qquad\qquad \times \frac{\Delta(z_1^2,\ldots,z_k^2) ^2
\prod_{j=1}^k z_j}
{\prod_{i=1}^k\prod_{j=1}^k(z_j-\alpha_i)(z_j+\alpha_i)}
e^{N\sum_{j=1}^kz_j} dz_1\cdots dz_k.
\end{eqnarray}


\subsection{Comparison with $L$-functions}

Note that
$\Lambda_M(s)=\prod_{n=1}^N(1-e^{i\theta_n}s)(1-e^{-i\theta_n}s)$
satisfies the functional equation $\Lambda_M(s)=s^{2N}
\overline{\Lambda_M}(1/s)$. However,  we can instead define

\begin{eqnarray}
\mathcal{Z}_M(s)&=&s^{-N}\Lambda_M(s),
\end{eqnarray}

\noindent which satisfies
$\mathcal{Z}_M(s)=\overline{\mathcal{Z}_M}(1/s)$, where
$\overline{\mathcal{Z}_M}(z) = \overline{\mathcal{Z}_M
(\overline{z})}$ and $\overline{z}$ denotes the complex conjugate
of~$z$ .

In \cite{kn:cfkrs} we conjecture the form of autocorrelation
functions of $L$-functions averaged over the family comprised of
$L(s,\chi_d)$, with $d$ a fundamental discriminant and
$\chi_d(n)=(\tfrac{d}{n})$, where here the family is ordered by
the conductor $d$.  In that paper the conjecture is formulated in
terms of a ``$Z$-function'' closely related to the $L$-function
but satisfying the functional equation
\begin{equation}
Z_L(s)= \overline{Z_L}(1-s).
\end{equation}
This is analogous to the random matrix function $\mathcal{Z}_M(s)$
and its functional equation, because the transformation from $s$
to $1-s$ in the number theory case reflects round the symmetry
point of the zeros of the $L$-function in the same manner as the
transformation from $s$ to $1/s$ in the random matrix theory case
reflects around the symmetry point of the eigenvalues.

The family of $L$-functions just defined is said to show symplectic symmetry
\cite{kn:katzsarnak99a,
kn:rub98} in as much as the statistics of the zeros around the symmetry point
are those of the eigenvalues of random matrices from $USp(2N)$.

The conjecture stated in \cite{kn:cfkrs} is then

\begin{conjecture}
\label{conj:Sp}

Suppose $g(u)$ is a suitable weight function. Then, if
$\mathcal{F}$ is the family of real Dirichlet $L$-functions with
fundamental discriminants $d<0$ (the sum over these fundamental
discriminants is indicated by $\sum^*$) we have
\begin{eqnarray}
&&\hspace{-0.3 in}\sum_{L\in{\mathcal{F}}} Z_L({\textstyle
\frac12}+\alpha_1)\cdots
                    Z_L({\textstyle \frac12}+\alpha_{k})g(|d|)
                 =   \sum^{\hspace{0.2 in}*}_{d<0}\,Q_k(\alpha,\log
\tfrac{|d|}{ 2\pi}) g(|d|)(1+O(|d|^{-\tfrac12 + \epsilon})) ,
\end{eqnarray}
\noindent in which
\begin{eqnarray}
\label{eq:Q}
 Q_k(\alpha, x)&=& \frac{(-1)^{k(k-1)/2}2^k}{k!}
\frac{1}{(2\pi
i)^{k}} \\
& &\quad \times \oint \cdots \oint \frac{G_-(z_1,
\dots,z_{k})\Delta(z_1^2,\dots,z_{k}^2)^2 \prod_{j=1}^{k} z_j}
{\displaystyle \prod_{\ell =1}^{k} \prod_{j=1}^{k} (z_j -
\alpha_\ell)(z_j+\alpha_{\ell}) } e^{\tfrac
x2\sum_{j=1}^{k}z_j}~dz_1\dots dz_{k}, \nonumber
\end{eqnarray}
\noindent where the path of integration encloses the $\pm \alpha$'s.
Here
\begin{equation}
G_-(z_1,\dots,z_k)=A_k(z_1,\dots,z_k) \prod_{j=1}^k \left(
    \frac{\Gamma(\tfrac 34+\tfrac {z_j}2) 2^{z_j}}
         {\Gamma(\tfrac 34- \tfrac {z_j}2)}
\right)^{\tfrac 12} \prod_{1\le i\le j\le k}\zeta(1+z_i+z_j),
\end{equation}
\noindent and $A_k$ is the Euler product, which is absolutely convergent for
$|\Re z_j|<1/2$, for $j=1,\ldots,k$, defined by
\begin{eqnarray}
A_k(z_1,\dots,z_k)& = &\prod_p \prod_{1\le i \le j \le k}
\left(1-\frac{1}{p^{1+z_i+z_j}}\right) \\
& &  \times \left(\frac
12 \left(\prod_{j=1}^k\left( 1-\frac{1}{p^{\tfrac
12+z_j}}\right)^{-1} +
\prod_{j=1}^k\left(1+\frac{1}{p^{\tfrac12+z_j}}\right)^{-1}
\right)+\frac 1p \right) \left( 1+ \frac{1}{p}\right)^{-1}. \nonumber
\end{eqnarray}

There is a similar
conjecture for the analogous sum over positive fundamental
discriminants. For this conjecture $G_-$ is replaced by $G_+$,
where
\begin{equation}
G_+(z_1,\dots,z_k)=A_k(z_1,\dots,z_k) \prod_{j=1}^k \left(
    \frac{\Gamma(\tfrac 14+\tfrac {z_j}2) 2^{z_j}}
         {\Gamma(\tfrac 14- \tfrac {z_j}2)}
\right)^{\tfrac 12} \prod_{1\le i\le j\le k}\zeta(1+z_i+z_j),
\end{equation}
\noindent and $A_k$ is as before.

\end{conjecture}

When comparing $\int_{USp(2N)}\mathcal{Z}_M(e^{-\alpha_1}) \cdots
\mathcal{Z}_M(e^{-\alpha_k}) dM$, which is very closely related to
(\ref{eq:SpBH}), with the autocorrelation function (\ref{eq:Q}) in
Conjecture~\ref{conj:Sp}, we note that equating the density of
zeros gives an equivalence $N=\tfrac{1}{2} \log
\tfrac{|d|}{2\pi}$.  Then we see immediately that the structure of
the $k$-fold integrals is very similar. The role of $\prod_{1\leq
\ell \leq m \leq k} (1-e^{-z_m-z_\ell})^{-1}$ in (\ref{eq:SpBH})
is played by $G_{\pm}(z_1,\ldots,z_k)$ in the $L$-function case.
Note that in both cases this factor produces poles when
$z_m=-z_{\ell}$, for $1\leq \ell \leq m \leq k$.  Extra arithmetic
information is in evidence in the $A_k$ factor in $G_{\pm}$ which,
of course, does not feature in the random matrix result.  Again,
the underlying similarity between the two formulae lends support
to the number theoretical conjecture and illustrates the strong
connection between $L$-functions and random matrix theory.


\section{Orthogonal Group: $O(2N)$}

We now turn our attention to the group $O(2N)$ of $2N\times 2N$
orthogonal matrices.  This group divides into two halves: the
group $SO(2N)$ of matrices from $O(2N)$ with determinant +1, and
$O^-(2N)$ which is comprised of the matrices with determinant -1.
We will examine these two components separately.


\subsection{$O^-(2N)$}

We are considering orthogonal $2N\times 2N$ matrices with
determinant $-1$.  These matrices have eigenvalues at
$1,-1,e^{i\theta_1},e^{-i\theta_1},\ldots,e^{i\theta_{N-1}},e^{-i\theta_{N-1}}$.
The measure may be expressed in the form
\begin{eqnarray}
\hspace{-0.2 in}\frac{(-1)^{((N-1)^2-(N-1))/2}} {(N-1)!\; \pi^{N-1} 2^{2(N-1)}}
 \Delta(e^{i\theta_1},\ldots,e^{i\theta_N},e^{-i\theta_1},\ldots,
 e^{-i\theta_N}) \prod_{j=1}^{N-1} (e^{i\theta_j}-e^{-i\theta_j}) d\theta_1\cdots
d\theta_{N-1}.
\end{eqnarray}
The characteristic polynomial for one of these matrices can be defined as
\begin{equation}
\Lambda_M(s)=\det(I-Ms)=(1-s)(1+s) \prod_{n=1}^{N-1}
(1-e^{i\theta_n}s)(1-e^{-i\theta_n}s) .
\end{equation}
\noindent The autocorrelation function is then
\begin{eqnarray}
&&\hspace{-0.2 in}I(O^-(2N),w_1,\ldots,w_k)\equiv\int_{O^-(2N)}(-1)^k\Lambda_M(w_1)
\cdots \Lambda_M(w_k) dM\\
&&\;\;= \left( \frac{(-1)^{((N-1)^2-(N-1))/2}} {(N-1)!\; \pi^{N-1}
2^{2(N-1)}} \int_{0}^{2\pi} \cdots \int_{0}^{2\pi} d\theta_1
\cdots d
\theta_{N-1}\Delta(e^{i\theta_1},\ldots,e^{i\theta_N},e^{-i\theta_1},\ldots,
 e^{-i\theta_N}) \right.\nonumber \\
&&\qquad\qquad\left.\times \prod_{m=1}^k \prod_{n=1}^{N-1}
(e^{i\theta_n}-w_m) (e^{-i\theta_n}-w_m)\prod_{j=1}^{N-1}
(e^{i\theta_j} - e^{-i\theta_j}) \right) \times \prod_{m=1}^k
(w_m^2-1).\nonumber
\end{eqnarray}

Following exactly the calculation in the previous section for the
group $USp(2N)$,
\begin{eqnarray}
&&I(O^-(2N),w_1,\ldots,w_k)\nonumber \\
&&\qquad\qquad=\frac{\prod_{m=1}^k (w_m^2-1)} {\prod_{1\leq i<j\leq k}
(w_j-w_i)}\left[ \sum_{{0\leq i_1 <i_2 <\cdots <i_k\leq
2N+k-1}\atop{ i_j\equiv j-1 mod 2}} \left|
\begin{array}{ccc} w_1^{i_1}& \cdots & w_1^{i_k} \\ \vdots&
\ddots& \vdots \\ w_k^{i_1} &\cdots & w_k^{i_k} \end{array}
\right| \right].
\end{eqnarray}
This then leads to
\begin{eqnarray}
&&\hspace{-0.3 in}I(O^-(2N),w_1,\ldots,w_k) \\
 &&=\left(\prod_{m=1}^k (w_m^2-1)\right)\times w_1^{N-1}\cdots
w_k^{N-1}\left[ \sum_{\epsilon_j\in \{-1,1\}} (\prod_{j=1}^k
w_j^{\epsilon_j(N-1)}) \prod_{1\leq i\leq j \leq k} (1-
w_i^{-\epsilon_i}w_j^{-\epsilon_j})^{-1} \right].\nonumber
\end{eqnarray}
However, in terms where $\epsilon_m=1$, we have a factor
\begin{equation}
\frac{w_m^2-1}{1-w_m^{-2\epsilon_m}}=\frac{w_m^2-1}{1-w_m^{-2}}=w_m^2=\epsilon_m
w_m\times w_m^{\epsilon_m} ,
\end{equation}
\noindent and if $\epsilon_m=-1$, then
\begin{equation}
\frac{w_m^2-1}{1-w_m^{-2\epsilon_m}}=\frac{w_m^2-1}{1-w_m^2}=-1=
\epsilon_mw_m\times w_m^{\epsilon_m}.
\end{equation}
Therefore,
\begin{eqnarray}
&&I(O^-(2N),w_1,\ldots,w_k)\nonumber \\
&&\qquad \qquad=w_1^N\cdots w_k^N \left[
\sum_{\epsilon_j\in\{-1,1\}}(\prod_{j=1}^k \epsilon_j
w_j^{\epsilon_j N}) \prod_{1\leq i<j\leq k}
(1-w_i^{-\epsilon_i}w_j^{-\epsilon_j})^{-1} \right],
\end{eqnarray}
\noindent and
\begin{eqnarray}
I(O^-(2N),e^{\alpha_1},\ldots,e^{\alpha_k})&=&\frac{(-1)^{k(k-1)/2}
2^k} {(2\pi i)^k k!}  e^{N\sum_{j=1}^k \alpha_j} \oint \cdots
\oint \prod_{1\leq \ell
\leq m \leq k} (1-e^{-z_m-z_\ell})^{-1} \nonumber \\
&& \;\;\;\; \times \frac{\Delta(z_1^2,\ldots,z_k^2) ^2
\prod_{j=1}^k \alpha_j}
{\prod_{i=1}^k\prod_{j=1}^k(z_j-\alpha_i)(z_j+\alpha_i)}
e^{N\sum_{j=1}^kz_j} dz_1\cdots dz_k,
\end{eqnarray}
\noindent using Lemma \ref{lemmaint}.


\subsection{$SO(2N)$}
\label{sect:SO2N}

We now consider the group of $2N \times 2N$ orthogonal matrices
which have positive determinant.  The eigenvalues of such matrices
come in complex conjugate pairs
$e^{i\theta_1},e^{-i\theta_1},\ldots,
e^{i\theta_N},e^{-i\theta_N}$.  The measure is
\begin{eqnarray}
\hspace{-0.2 in}\frac{(-1)^{N(N-1)/2}2^{-2N+1}} {\pi^N N!}
\Delta(e^{i\theta_1},\ldots,e^{i\theta_N}
,e^{-i\theta_1},\ldots,e^{-i\theta_N}) \prod_{k=1}^N
(e^{-i\theta_k}-e^{i\theta_k})^{-1}d\theta_1\cdots d\theta_N.
\end{eqnarray}
The characteristic polynomial for these matrices is
\begin{equation}
\Lambda_M(s)=\det(I-Ms)=\prod_{n=1}^N (1-e^{i\theta_n}s)(1-e^{-i\theta_n}s),
\end{equation}
\noindent so the autocorrelation function which we wish to
evaluate is
\begin{eqnarray}
&&\hspace{-0.3 in}I(SO(2N),w_1,\ldots,w_k)\equiv \int_{SO(2N)} \Lambda_M(w_1)
\cdots \Lambda_M(w_k) dM \nonumber\\
&& =\frac{(-1)^{N(N-1)/2}2^{-2N+1}} {\pi^N N!}\int_{0}^{2\pi}
\cdots \int_{0}^{2\pi}
\Delta(e^{i\theta_1},\ldots,e^{i\theta_N},e^{-i\theta_1},\ldots,
e^{-i\theta_N}) \nonumber \\
&&\qquad\times \prod_{m=1}^k\prod_{n=1}^{N} (e^{i\theta_n}
-w_m) (e^{-i\theta_n}-w_m) \left[\prod_{n=1}^N
(e^{-i\theta_n}-e^{i\theta_n})\right]^{-1} d\theta_1 \cdots d\theta_N\nonumber \\
&&=\frac{(-1)^{N(N-1)/2}2^{-2N+1}} {\pi^N N!}\frac{1}{\prod_{1\leq
i<j\leq k} (w_j-w_i)} \int_{0}^{2\pi} \cdots \int_{0}^{2\pi}
\left[\prod_{n=1}^N(e^{-i\theta_n}-e^{i\theta_n})\right]^{-1} \nonumber \\
&&\qquad\times \Delta(w_1,\ldots,w_k,e^{i\theta_1},\ldots,
e^{i\theta_N},e^{-i\theta_1},\ldots,e^{-i\theta_N}) d\theta_1\cdots d\theta_N\nonumber\\
&&=\frac{2^{-2N+1}} {\pi^N N!} \frac{1}{\prod_{1\leq i<j\leq k}
(w_j-w_i)} \int_{0}^{2\pi} \cdots \int_{0}^{2\pi}
\left[\prod_{n=1}^N(e^{-i\theta_n}-e^{i\theta_n})\right]^{-1}\nonumber \\
&&\qquad \times \sum_{\sigma\in S_{2N+k}} {\rm sgn} \sigma
\;w_1^{\sigma(1)-1} w_2^{\sigma(2)-1} \cdots w_k^{\sigma(k)-1}
e^{i[\sigma(k+1)-1]\theta_1} e^{-i[\sigma(k+2)-1]\theta_1}
\cdots\label{eq:preintegration}\\
&&\qquad\qquad\qquad\qquad\qquad e^{i[\sigma(k+2N-1)-1]\theta_N}
e^{-i[\sigma(k+2N)-1]\theta_N}d\theta_1\cdots
d\theta_N,\nonumber
\end{eqnarray}
\noindent where in the final line we have the determinant
expansion of $\Delta(w_1,\ldots,w_k,e^{i\theta_1},e^{-i\theta_1},
\ldots,$ $ e^{i\theta_N},e^{-i\theta_N})=(-1)^{N(N-1)/2}
\Delta(w_1,\ldots,w_k,e^{i\theta_1},\ldots,
e^{i\theta_N},e^{-i\theta_1},\ldots,e^{-i\theta_N})$ expressed in
terms of the permutations of $\{1,2,\ldots,2N+k\}$ .

The sum over $\sigma\in S_{2N+k}$ in (\ref{eq:preintegration}) can
be broken up and written as follows
\begin{eqnarray}
&&\sum_{\delta\in D} {\rm sgn}(\delta) \left( \sum_{\alpha\in A}
{\rm sgn}(\alpha) w_1^{\alpha(1)-1} \cdots w_k^{\alpha(k)-1}
\right) \nonumber \\
&&\qquad \times\left( \sum_{\beta \in B} {\rm sgn}(\beta) \prod_{j=1}^N (
e^{i\theta_j(\beta(2j-1)-\beta(2j))}-e^{-i\theta_j(\beta(2j-1)-\beta(2j))})\right)
,\label{eq:subterms}
\end{eqnarray}
\noindent where $D\subset S_{2N+k}$ is the set of permutations
such that $\delta(1)<\cdots<\delta(k)$ and $\delta(k+1)<\cdots
<\delta(k+2N)$, $A$ is the set of all permutations of
$\delta(1),\ldots,\delta(k)$, and $B$ is the set of permutations
of $\delta(k+1),\ldots,\delta(k+2N)$ such that
$\beta(1)<\beta(2)$,$\beta(3)<\beta(4)$,$\ldots$,$\beta(2N-1)<\beta(2N)$.

 Using $x^N-y^N=(x-y)(x^{N-1}+x^{N-2}y+\cdots + x
y^{N-2} +y^{N-1})$, we see that the product over $j$ in
(\ref{eq:subterms}) contains a factor $\prod_{n=1}^N
(e^{-i\theta_n}-e^{i\theta_n})$ which cancels with the identical
factor in (\ref{eq:preintegration}).  Since the integral in
(\ref{eq:preintegration}) integrates to zero unless the integrand
is independent of all $\theta_j$, and since in our case
$x=e^{-i\theta}$ and $y=e^{i\theta}$,  we obtain zero for any term
in the sum over $\beta$  unless $\beta(2j) -\beta(2j-1)$ is an odd
number for all $j=1,2,\ldots,N$, in which case
(\ref{eq:preintegration}) reduces to
\begin{eqnarray}
&&\frac{2^{-N+1}} {N!} \frac{1}{\prod_{1\leq i<j\leq k}
(w_j-w_i)}\sum_{\delta\in D} {\rm sgn}(\delta) \left(
\sum_{\alpha\in A} {\rm sgn}(\alpha) w_1^{\alpha(1)-1} \cdots
w_k^{\alpha(k)-1} \right)\nonumber \\
&&\qquad \qquad\qquad\times\left( \sum_{{\beta \in B}\atop
{\beta(2j) -\beta(2j-1) = {\rm odd}}} {\rm sgn}(\beta) \right)
.\label{eq:subterms2}
\end{eqnarray}
To perform the remaining sum over $\beta$, recall that the
permutation $\beta$ rearranges $\delta(k+1),\ldots,\delta(2N+k)$
(which are arranged in ascending order), and note that the sum
over $\beta$ in (\ref{eq:subterms2}) will contain zero terms
unless $N$ of $\delta(k+1),\ldots,\delta(2N+k)$ are even and $N$
are odd.  In particular, one of $\beta(2j-1)$ and $\beta(2j)$ must
be even and one must be odd for each $j=1,\ldots,N$.  To perform
the $\beta$ sum, we essentially need to count (with signs) all the
ways to pair up each even number with an odd number.  It can be seen
that if in the original ascending order $\delta(k+1),\ldots,\delta(2N+k)$
even and odd numbers alternate, then the sum over $\beta$ is given by
$N!$ times the $N\times N$ determinant
\begin{equation}
\left| \begin{array}{ccccc} 1& 1& 1& \cdots &1 \\ -1 & 1& 1 &
\cdots & 1 \\ -1&-1 & 1& \cdots &1\\ \vdots& \vdots &\vdots &
\ddots & \vdots \\ -1 & -1& -1& \cdots&1 \end{array} \right|=2^{N-1},
\end{equation}
where the determinant accounts (with sign) for the pairing of each even number
with an odd number, while the $N!$ accounts for the further permutation
of the $N$ pairs.  The same reasoning produces an $N\times N$ determinant which
is zero when the arrangement $\delta(k+1),\ldots,\delta(2N+k)$ contains
two consecutive even or two consecutive odd numbers.  Noting that
${\rm sgn} \delta$ in (\ref{eq:subterms}) is always +1 for
$\delta$ such that even and odd numbers alternate in
$\delta(k+1),\ldots,\delta(2N+k)$, we arrive at
\begin{eqnarray}
I(SO(2N),w_1,\ldots,w_k)&=& \frac{1}{\prod_{1\leq i<j\leq k}
(w_j-w_i)}\sum_{i_1,\ldots,i_k}\left|
\begin{array}{ccc} w_1^{i_1} & \cdots &
w_1^{i_k}\\ \vdots& \ddots & \vdots \\
w_k^{i_1} & \cdots & w_k^{i_k}
\end{array}\right| ,
\label{eq:SOdet}
\end{eqnarray}
\noindent where the conditions on $i_1,\ldots,i_k$ are that
$i_j\in\{0,1,\ldots,2N+k-1\}$, $i_1<i_2<\cdots<i_k$ and
\begin{eqnarray}
&k\;{\rm even}& \left\{ \begin{array}{c}
i_1=0,i_2=i_3-1,i_4=i_5-1,\ldots,i_{k-2}=i_{k-1}-1,i_k=2N+k-1 \\
{\rm or}\\
i_1=i_2-1,i_3=i_4-1,\ldots,i_{k-1}=i_k-1 \end{array} \right. \nonumber \\
&k\;{\rm odd}& \left\{\begin{array}{c}
i_1=0,i_2=i_3-1,i_4=i_5-1,\ldots,i_{k-1}=i_k-1\\
{\rm or} \\
i_1=i_2-1,i_3=i_4-1,\ldots,i_{k-2}=i_{k-1}-1,i_k=2N+k-1.
\end{array} \right.\label{eq:conditions}
\end{eqnarray}

Once more, this is a sum over Schur functions,
\begin{eqnarray}
&&I(SO(2N),w_1,\ldots,w_k)\nonumber\\
&&=\sum_{{\lambda' {\rm \;odd}}\atop{k\geq \lambda_1'\geq \cdots
\geq \lambda_{2N}'>0}} S_{\lambda}(w_1,\ldots,w_k)
+\sum_{{\lambda' {\rm \;even}}\atop{k\geq \lambda_1'\geq \cdots
\geq \lambda_{2N}'\geq0}} S_{\lambda}(w_1,\ldots,w_k)
\end{eqnarray}
where the sum is over partitions
$\lambda'=(\lambda_1',\ldots,\lambda_{2N}')$, with no part greater
than $k$, and $\lambda'$ is the conjugate partition to $\lambda$.
Note that the condition on the odd $\lambda'$ implies that the
partition has exactly $2N$ non-zero parts, whereas the sum over
even partitions only requires that $\lambda'$ has no more than
$2N$ non-zero parts.

We now show that (\ref{eq:SOdet}) may be expressed in the form
\begin{eqnarray}
\label{eq:SOmoment1}
&& I(SO(2N),w_1,\ldots,w_k)\nonumber \\
&&\qquad\qquad= w_1^N\cdots w_k^N \left[ \sum_{\epsilon_j\in \{-1,1\}}\left(
\prod_{j=1}^k w_j^{N\epsilon_j} \right) \prod_{1\leq i<j\leq k}
(1-w_i^{-\epsilon_i}w_j^{-\epsilon_j})^{-1} \right].
\end{eqnarray}

In order to prove (\ref{eq:SOmoment1}) we must first prove an identity
very similar in form to Identity \ref{iden:1} in the
symplectic symmetry section.  This is
\begin{identity}
\label{iden:4}
\begin{eqnarray}
&&\sum_{j=1}^n w_j^2 \Delta(w_1,\ldots,w_n)\big|
_{w_j=0} \prod_{m\neq j} (1-w_mw_j)\nonumber \\
&&\qquad\qquad=\left\{ \begin{array}{cc}  w_1^2\cdots w_n^2
\Delta(w_1,\ldots,w_n) & {\rm\;if\;} n {\rm \;odd} \\
(w_1^2\cdots w_n^2-w_1\cdots w_n) \Delta(w_1,\ldots,w_n) &
{\rm\;if\;}n{\rm\;even} \end{array} \right..\nonumber
\end{eqnarray}
\end{identity}

To prove this we rewrite the factor $w_j^2$ as $1-(1-w_j^2)$. The
contribution to Identity \ref{iden:4} from the $(1-w_j^2)$ term,
is just the right side of Identity \ref{iden:1}; that is,
$\Delta(w_1,\ldots,w_n)(1-w_1^2\cdots w_n^2)$. The remainder of
the left side of Identity \ref{iden:4} we write as in
(\ref{eq:fdet}), where $f(w_j)=\prod_{m\neq j} (1-w_mw_j)$.  Note
that we cannot immediately apply Lemma \ref{lemma1} because $f(x)$
is not symmetric amongst the $w$'s.  However, if we write
$f(w_j)=\prod_{m\neq j}(1-w_mw_j)=\sum_{i=0}^{n-1}a_iw_j^i$ and
$g(w)=\prod_{m=1}^n (1-w_mw)=\sum_{i=0}^nb_i w^i$, then the
identity $g(w_j)=f(w_j)(1-w_j^2)$ produces the recurrence relation
$b_0=a_0=1$, $b_i=a_i-w_ja_{i-1}$ (for $i=1,\ldots,n-1$) and
$b_n=-w_j a_{n-1}$.  This allows us to write
\begin{eqnarray}
\label{eq:symmb}
 \hspace{-0.2 in}\sum_{i=0}^{n-1} a_i w_j^i &=& \left\{
\begin{array}{cc} \sum_{i=0}^{(n-1)/2} \sum_{q=0}^{i} w_j^{2i-q}
b_{q} +\sum_{i=(n+1)/2}^{n-1}
\sum_{q=0}^{n-i-1} -w_j^{2i-n+q} b_{n-q}& n{\rm \; odd} \\
\sum_{i=0}^{n/2} \sum_{q=0}^{i} w_j^{2i-q} b_q +
\sum_{i=(n+2)/2}^{n-1} \sum_{q=0}^{n-i-1} -w_j^{2i-n+q} b_{n-q} &
n{\rm \; even} \end{array}.\right.
\end{eqnarray}

Since the $b$'s are symmetric functions of the $w$'s and in
(\ref{eq:fdet}) we have columns containing powers of the $w$'s
from 1 to $n-1$, the only terms in the expression for
$f(w_j)=\sum_{i=0}^{n-1} a_i w_j^i $ on the right side of
(\ref{eq:symmb}) above which {\em cannot} be cancelled by adding
or subtracting one of these columns multiplied by a symmetric
function of the $w$'s are $a_0=1$ and, in the case that $n$ is
even, $w_j^n$. The determinant is then easily evaluated as
$\Delta(w_1,\ldots,w_n)$ if $n$ is odd, and
$\Delta(w_1,\ldots,w_n)-w_1\ldots w_n \Delta(w_1,\ldots, w_n)$ if
$n$ is even.  This proves Identity \ref{iden:4}.

Now we move on to determining the form of the autocorrelation
functions in (\ref{eq:SOmoment1}).  When $k$ is even in
(\ref{eq:SOmoment1}) we will write
\begin{eqnarray}
I(SO(2N),w_1,\ldots,w_{2k})=I_{2N+2k-1}^M(w_1,\ldots,w_{2k})+I_{2N+2k-1}^E(w_1
,\ldots,w_{2k}),
\end{eqnarray}
\noindent where
\begin{equation}
I^M_n(w_1,\ldots,w_{2k})\equiv \frac{1}{\Delta(w_1,\ldots,w_{2k})}
\sum_{{i_1<\cdots<i_{2k}\in\{0,\ldots,n\}}\atop{i_{2j-1}=i_{2j}-1,j=1,\ldots,k}}
\left| \begin{array}{ccc}w_1^{i_1}& \cdots& w_1^{i_{2k}} \\ \vdots
& \ddots & \vdots \\ w_{2k}^{i_1} & \cdots & w_{2k}^{i_{2k}}
\end{array} \right| \label{eq:IMdef}
\end{equation}
\noindent and
\begin{equation}
I^E_n(w_1,\ldots,w_{2k})\equiv \frac{1}{\Delta(w_1,\ldots,w_{2k})}
\sum_{{i_1<\cdots<i_{2k}\in\{0,\ldots,n\}}\atop{i_1=0,i_{2j}=i_{2j+1}-1,j=1,
\ldots,k-1,i_{2k}=n}} \left| \begin{array}{ccc}w_1^{i_1}& \cdots&
w_1^{i_{2k}} \\ \vdots & \ddots & \vdots \\ w_{2k}^{i_1} & \cdots
& w_{2k}^{i_{2k}}
\end{array} \right|.\label{eq:IEdef}
\end{equation}

Similarly, when $k$ is odd in (\ref{eq:SOmoment1}) we will write
\begin{equation}
I(SO(2N),w_1,\ldots,w_{2k+1})=I^R_{2N+2k}(w_1,\ldots,w_{2k+1})+I^L_{2N+2k}
(w_1,\ldots,w_{2k+1}),
\end{equation}
\noindent where
\begin{equation}
I^R_n(w_1,\ldots,w_{2k+1})\equiv
\frac{1}{\Delta(w_1,\ldots,w_{2k+1})}
\sum_{{i_1<\cdots<i_{2k+1}\in\{0,\ldots,n\}}\atop{i_{2j-1}=i_{2j}-1,j=1,\ldots,k,
i_{2k+1}=n}} \left| \begin{array}{ccc}w_1^{i_1}& \cdots&
w_1^{i_{2k+1}} \\ \vdots & \ddots & \vdots \\ w_{2k+1}^{i_1} &
\cdots & w_{2k+1}^{i_{2k+1}}
\end{array} \right|\label{eq:IRdef}
\end{equation}
\noindent and
\begin{equation}
I^L_n(w_1,\ldots,w_{2k+1})\equiv
\frac{1}{\Delta(w_1,\ldots,w_{2k+1})}
\sum_{{i_1<\cdots<i_{2k+1}\in\{0,\ldots,n\}}\atop{i_1=0,i_{2j}=i_{2j+1}-1,j=1,
\ldots,k}} \left| \begin{array}{ccc}w_1^{i_1}& \cdots&
w_1^{i_{2k+1}} \\ \vdots & \ddots & \vdots \\ w_{2k+1}^{i_1} &
\cdots & w_{2k+1}^{i_{2k+1}}
\end{array} \right|. \label{eq:ILdef}
\end{equation}

We now prove the following identities
\begin{subequations}
\label{eq:Iconj}
\begin{eqnarray}
&&I^M_n(w_1,\ldots,w_{2k}) \label{eq:IMconj}\\
&&\qquad=\frac{1}{\mathcal{E}([2k])\Delta ([2k])}
\sum_{{A\cup B =[2k],\;A\cap B=\emptyset}\atop{|B|{\rm \;even}}}
w_A^n E(A,B)\Delta (A)\Delta (B)(-1)^
{S(A,B)-|A|}, \nonumber\\
&&I^E_n(w_1,\ldots,w_{2k}) \label{eq:IEconj}\\
&&\qquad =\frac{1}{\mathcal{E}([2k])\Delta ([2k])}
\sum_{{A\cup B =[2k],\;A\cap B=\emptyset}\atop{|B|{\rm \;odd}}}
w_A^n E(A,B)\Delta (A)\Delta (B)(-1)^
{S(A,B)-|A|}, \nonumber\\
&&I^R_n(w_1,\ldots,w_{2k+1})\label{eq:IRconj}\\
&&\qquad=\frac{1}{\mathcal{E}([2k+1])\Delta ([2k+1])}
\sum_{{A\cup B =[2k+1],\;A\cap B=\emptyset}\atop{|B|{\rm \;even}}}
w_A^n E(A,B)\Delta (A)\Delta (B)(-1)^
{S(A,B)-|A|}, \nonumber
\end{eqnarray}
and
\begin{eqnarray}
&&I^L_n(w_1,\ldots,w_{2k+1})\label{eq:ILconj}\\
&& \qquad =\frac{1}{\mathcal{E}([2k+1])\Delta ([2k+1])}
\sum_{{A\cup B =[2k+1],\;A\cap B=\emptyset}\atop{|B|{\rm \;odd}}}
w_A^n E(A,B)\Delta (A)\Delta (B)(-1)^ {S(A,B)-|A|}.\nonumber
\end{eqnarray}
\end{subequations}
\noindent Here the only notation not already defined in Section
\ref{sect:Sp} is
\begin{eqnarray}
\mathcal{E}(A)&=&\prod_{{m<n}\atop {m,n\in A}}(1-w_mw_n).
\end{eqnarray}

For the case $k=1$ it is easy to show that (\ref{eq:Iconj}) holds.
We now prove (\ref{eq:Iconj}) for any $k$ by induction.
First we note that Identity \ref{iden:4} can be written as
\begin{eqnarray}
\label{eq:iden3revisited} &&\hspace{-0.5 in}\sum_{j\in A} w_j \Delta (A_j) E(\{j\},A_j)
(-1)^{W(A_j,\{j\})} = (-1)^{|A|-1} \Delta (A) \left\{ \begin{array}{cc}
(w_A-1)& {\rm \;if\;}|A|{\rm\;even} \\ w_A& {\rm
\;if\;}|A|{\rm\;odd}\end{array}\right.,
\end{eqnarray}
\noindent where $A\subset\{1,2,\ldots,m\}\equiv [m]$, $A_j$ is the
set of elements of $A$ with $j$ removed, and $|A|$ is the number
of elements in the set $A$.  Also, $w_A=\prod_{m\in A}w_m$ and
$W(A,B)=\sum_{ {m\in A,\;n\in B}\atop {m>n}} 1$.

If we make use of Identity \ref{iden:3} with $x=1$, then in the
current notation this appears as
\begin{eqnarray}
\label{eq:iden4revisited} && \sum_{{A\cup B=[m],A\cap B=\emptyset}
\atop {|A| {\rm \;even}}}(-1)^{S(A,B)} E(A,B)\Delta (A)\Delta
(B)w_A^{m-2}=0.
\end{eqnarray}

To prove the form of $I^R$ in (\ref{eq:IRconj}), we need to show
that
\begin{eqnarray}
\label{eq:prove1}
&&\sum_{{i_1<\cdots<i_{2k+1}\in\{0,\ldots,n\}}\atop{i_{2j-1}=i_{2j}-1,j=1,\ldots,k,
i_{2k+1}=n}} \left| \begin{array}{ccc}w_1^{i_1}& \cdots&
w_1^{i_{2k+1}} \\ \vdots & \ddots & \vdots \\ w_{2k+1}^{i_1} &
\cdots & w_{2k+1}^{i_{2k+1}}
\end{array} \right|\nonumber \\
&&\qquad=\frac{1}{\mathcal{E}([2k+1])} \sum_{{A\cup B =[2k+1],\;A\cap
B=\emptyset}\atop{|B|{\rm \;even}}} w_A^n E(A,B)\Delta (A)\Delta (B)(-1)^
{S(A,B)-|A|}.
\end{eqnarray}
Using the definition of $I^M$ (\ref{eq:IMdef}), we see that the
left side of the above is
\begin{eqnarray}
&&\sum_{j=1}^{2k+1} (-1)^{j-1}
w_j^n\Delta ([2k+1]_j)I_{n-1}^M([2k+1]_j),
\end{eqnarray}
\noindent and then by induction using (\ref{eq:IMconj}), the line
above equals
\begin{eqnarray}
&& \hspace{-0.2 in}\sum_{j=1}^{2k+1} (-1)^{j-1} w_j^n
\frac{1}{\mathcal{E}([2k+1]_j)} \sum _{{F\cup B=[2k+1]_j,\;F\cap
B=\emptyset}\atop {|B|{\rm\;even}}} w_F^{n-1} E(F,B) \Delta
(F)\Delta (B) (-1)^{S(F,B)-|F|}.
\end{eqnarray}
Now we define $A=F\cup \{j\}$ and then exchange the order of the
two sums, to obtain
\begin{eqnarray}
&& \sum_{{A\cup B = [2k+1],\;A\cap B=\emptyset}\atop
{|B|{\rm\;even}}} \sum_{j\in A} \frac{(-1)^{j-1}
w_j^n}{\mathcal{E}([2k+1]_j)} (-1)^{S(A_j,B)-|A_j|} w_{A_j}^{n-1}
E(A_j,B)\Delta (A_j)\Delta (B).
\end{eqnarray}

We note that $\frac{E(A_j,B)}{\mathcal{E}(A_j\cup B)} =
\frac{E(A,B) E(\{j\},A_j)} {\mathcal{E}(A\cup B)}$ and that if
$A\cup B=[m]$, then $(-1)^{
j+S(A_j,B)-|A_j|}=(-1)^{S(A,B)-W(A_j,\{j\})}$ if $m$ is odd, and
$(-1)^{ j+S(A_j,B)-|A_j|}$ $=(-1)^{S(A,B)-W(A_j,\{j\})-1}$ if $m$
is even.
So we have
\begin{eqnarray}
&&\hspace{-.8 in} \sum_{{A\cup B = [2k+1],\;A\cap
B=\emptyset}\atop {|B|{\rm\;even}}} w_A^{n-1}
\frac{E(A,B)}{\mathcal{E}([2k+1])} (-1)^{S(A,B)} \Delta (B)
\sum_{j\in A} (-1) ^{W(A_j,\{j\})-1} w_j E(\{j\},A_j)\Delta (A_j).
\end{eqnarray}
Using (\ref{eq:iden3revisited}), we then find
\begin{equation}
\frac{1}{\mathcal{E}([2k+1])}\sum_{{A\cup B=[2k+1],\;A\cap
B=\emptyset} \atop{|B|{\rm \;even}}}w_A^n E(A,B)\Delta (A)\Delta (B)
(-1)^{S(A,B)-|A|},
\end{equation}
\noindent which proves (\ref{eq:prove1}) and so confirms the form
of $I^R$ in (\ref{eq:IRconj}).

The expression for $I^M$ is proved similarly, using induction and the
 form of $I^R$ found in (\ref{eq:IRconj}).  We need to
show that
\begin{eqnarray}
\label{eq:prove2} &&\sum_{{i_1<\cdots<i_{2k}\in \{0,\ldots,n\}}
\atop {i_{2j-1}=i_{2j}-1,j=1,\ldots ,k}}\left|
\begin{array}{ccc}w_1^{i_1}& \cdots& w_1^{i_{2k}} \\ \vdots &
\ddots & \vdots \\ w_{2k}^{i_1} & \cdots & w_{2k}^{i_{2k}}
\end{array} \right|\nonumber \\
&&\qquad\qquad=\frac{1}{\mathcal{E}([2k])} \sum_{{A\cup B=[2k],A\cap
B=\emptyset} \atop {|B| {\rm \; even}}} w_A^n
E(A,B)\Delta (A)\Delta (B)(-1)^{S(A,B)-|A|}.
\end{eqnarray}
The left side of this expression, written in terms of $I^R$, is
\begin{eqnarray}
\label{eq:withIR} && \sum_{i_{2k}=2k-1}^{n} \sum_{j=1}^{2k} (-1)^j
w_j^{i_{2k}} \Delta ([2k]_j) I_{i_{2k}-1}^R([2k]_j).
\end{eqnarray}

Now we proceed by induction and use (\ref{eq:IRconj}).
Continuing exactly as we did in the case of $I^R$ above, we find that
(\ref{eq:withIR}) reduces to
\begin{eqnarray}
&& \sum_{{A\cup B=[2k],A\cap B=\emptyset} \atop {|B|{\rm
\;even}}} \frac{E(A,B)}{\mathcal{E}([2k])} (-1)^{S(A,B)}(
w_A^n-w_A^{2k-2}) \Delta (A)\Delta (B).
\end{eqnarray}
Finally, by identity (\ref{eq:iden4revisited}), we see that all
the terms in which $w_A$ appears with exponent $2k-2$ disappear,
leaving us with
\begin{equation}
\sum_{{A\cup B=[2k],A\cap B=\emptyset} \atop {|B|{\rm \;even}}}
\frac{E(A,B)}{\mathcal{E}([2k])} (-1)^{S(A,B)}w_A^n \Delta (A)\Delta (B),
\end{equation}
\noindent which proves (\ref{eq:prove2}) and so also (\ref{eq:IMconj}).

To prove (\ref{eq:IEconj}) for $I^E$ and (\ref{eq:ILconj})
for $I_L$, we follow exactly the same procedure as above.

Finally, we show that the form of the expressions in
(\ref{eq:Iconj}) can be written as sums over
$\epsilon_j\in\{-1,1\}$ and so complete the proof of
(\ref{eq:SOmoment1}). Note that, using $D(A,B)=(-1)^{W(A,B)}\Delta
([m])/(\Delta (A)\Delta (B))$ and $\mathcal{E}
(A)\mathcal{E}(B)=\mathcal{E}([m])/E(A,B)$ (with $A\cup
B=[m]\equiv \{1,\ldots ,m\}$, $A\cap B=\emptyset$) and letting
``parity'' stand for {\em either} ``even'' or ``odd'',
\begin{eqnarray}
&&\frac{1}{\mathcal{E}([m])\Delta ([m])} \sum_{{A\cup B=[m],A\cap
B=\emptyset}
\atop {|B| {\rm \;parity}}} w_A^n E(A,B)\Delta (A)\Delta (B) (-1)^{S(A,B)-|A|} \nonumber\\
&&\qquad\qquad= \sum_{{A\cup B=[m],A\cap B=\emptyset} \atop {|B| {\rm
\;parity}}} w_A^{n-m+1} \frac{(-1)^{|A|(|A|-1)/2} w_A^{|A|-1}
}{\mathcal{E}(A)} \frac{1}{\mathcal{E}(B)} \frac{(-1)^{|A||B|}
w_A^{|B|}} {D(A,B)} \nonumber
\end{eqnarray}
\begin{eqnarray}
&&=w^{\tfrac{n-m+1}{2}}\sum_{{A\cup B=[m],A\cap B=\emptyset} \atop
{|B| {\rm \;parity}}} w_A^{\tfrac{n-m+1}{2}}
w_B^{-\tfrac{n-m+1}{2}} \left( \prod_{{p,q\in A}\atop{p<q}}
\frac{1}{1-\tfrac{1}{w_pw_q}} \right) \nonumber \\
&&\qquad \qquad \times\left( \prod_{{p,q\in B}\atop{p<q}}
\frac{1}{1-w_pw_q}\right)
\left( \prod_{{p\in A}\atop{q\in B}} \frac{1}{1-\tfrac{w_q}{w_p}}\right) \nonumber \\
&&= w^{\tfrac{n-m+1}{2}} \sum_{{\epsilon_j\in\{1,-1\}}\atop
{\prod_j \epsilon_j =(-1)^{\rm parity}}} \left(\prod_{\ell=1}^{m}
w_{\ell}^{\epsilon_{\ell}(\tfrac{n-m+1} {2})}\right)\left(
\prod_{1\leq q<\ell\leq
m}(1-w_q^{-\epsilon_q}w_{\ell}^{-\epsilon_{\ell}})^{-1}\right).
\end{eqnarray}
Thus we end up with
\begin{equation}
I(SO(2N),w_1,\ldots,w_k)=w_1^N\cdots
w_k^N\left[\sum_{\epsilon_j\in \{1,-1\}} \left(
\prod_{j=1}^kw_j^{N\epsilon_j}\right) \prod_{1\leq i<j\leq k}
(1-w_i^{-\epsilon_i}w_j^{-\epsilon_j})^{-1}\right],
\end{equation}
\noindent which is exactly (\ref{eq:SOmoment1}).

Using Lemma \ref{lemmaint} we can express this as a multiple integral:

\begin{eqnarray}
&&I(SO(2N),e^{\alpha_1},\ldots,e^{\alpha_k}) \nonumber \\
&&\qquad\qquad=\frac{(-1)^{k(k-1)/2} 2^k} {(2\pi i)^k k!} e^{N\sum_{j=1}^k \alpha_j}\oint \cdots \oint
\prod_{1\leq \ell < m \leq k}(1-e^{-z_m-z_{\ell}})^{-1} \nonumber \\
&& \qquad \qquad\qquad\qquad\times\frac{\Delta(z_1^2,\ldots,z_k^2) ^2
\prod_{j=1}^k z_j}
{\prod_{i=1}^k\prod_{j=1}^k(z_j-\alpha_i)(z_j+\alpha_i)}\;
e^{N\sum_{j=1}^kz_j} dz_1\cdots dz_k.
\end{eqnarray}


\subsection{Comparison with $L$-functions}

In \cite{kn:cfkrs} we give a conjecture for the autocorrelation functions of
\begin{equation}
L_f(s)=\sum_{n=1}^{\infty} \lambda_f(n)n^{-s},
\end{equation}
\noindent near the critical point $s=1/2$ averaged over $f\in
H_k(N)$.  Here we denote by $H_k(N)$ the set of
primitive newforms $f\in S_k(\lambda_0(N))$ and the $\lambda_f$
are the Fourier coefficients of the newform. For simplicity, we
restrict attention to $k=2$ and $N=q$, a prime.  The zeros of
this family near the critical point display
orthogonal symmetry.

The $L$-function satisfies the functional equation
\begin{equation}
L_f(s)=\varepsilon_f X(s)L_f(1-s),
\end{equation}
\noindent with $\varepsilon_f=-\sqrt{q}\lambda_f(q) = \pm 1$. If
instead we define

\begin{equation}
Z_f(s)=X(s)^{-1/2}L_f(s),
\end{equation}
\noindent then $Z_f(s)$ obeys the functional equation
\begin{equation}
\label{eq:functionalZf}
 Z_f(s)=\varepsilon_f Z_f(1-s).
\end{equation}
After defining the ``harmonic average''
\begin{equation}
\sum_{f\in H_2(q)}^{\;\;\;\;\;\;\;\;\;\;\;\;h}
Z_f(1/2+\alpha_1)\cdots Z_f(1/2+\alpha_k)\equiv \sum_{f\in
H_2(q)}Z_f(1/2+\alpha_1)\cdots Z_f(1/2+\alpha_k) /<f,f>,
\end{equation}
\noindent we have the following three conjectures:

\begin{conjecture}
\label{conj:fullL}

\begin{eqnarray*}
 &&\hspace{-0.2 in}\sum_{f\in
H_{2}^*(q)}^{\;\;\;\;\;\;\;\;\;\;\;\;h}
Z_f(1/2+\alpha_1)\cdots
Z_f(1/2+\alpha_k)\\
&&\qquad=\sum_{\begin{array}{c} \epsilon_j\in \{-1,+1\}\nonumber\\
\prod_{j=1}^k \epsilon_j=1\end{array}} \prod_{j=1}^k
X(1/2+\epsilon_j\alpha_j)^{-1/2} \prod_{1\leq i<j\leq k}
\zeta(1+\epsilon_i\alpha_i+\epsilon_j\alpha_j)
A(\epsilon_1\alpha_1,\ldots,\epsilon_k\alpha_k)  \nonumber\\
&&\qquad \qquad\qquad\times (1+O(q^{-\tfrac{1}{2}+\epsilon})),\nonumber
\end{eqnarray*}

\end{conjecture}

\begin{conjecture}
\label{conj:evenL}

\begin{eqnarray}
 &&\hspace{-0.2 in}\sum_{\begin{array}{c}f\in H_{2}^*(q)\\f\;{\rm
even}\end{array}}^{\;\;\;\;\;\;\;\;\;\;\;\;h}
Z_f(1/2+\alpha_1)\cdots
Z_f(1/2+\alpha_k)\nonumber \\
&&\qquad=\frac{1}{2}\sum_{\epsilon_j\in \{-1,+1\}}\prod_{j=1}^k
X(1/2+\epsilon_j\alpha_j)^{-1/2} \prod_{1\leq i<j\leq k}
\zeta(1+\epsilon_i\alpha_i+\epsilon_j\alpha_j)
A(\epsilon_1\alpha_1,\ldots,\epsilon_k\alpha_k)  \nonumber\\
&&\qquad \qquad \qquad\times(1+O(q^{-\tfrac{1}{2}+\epsilon})) \nonumber
\end{eqnarray}
\end{conjecture}

\noindent and

\begin{conjecture}
\label{conj:oddL}

\begin{eqnarray}
 &&\hspace{-0.2 in}\sum_{\begin{array}{c}f\in H_{2}^*(q)\\f\;{\rm
odd}\end{array}}^{\;\;\;\;\;\;\;\;\;\;\;\;h}
Z_f(1/2+\alpha_1)\cdots
Z_f(1/2+\alpha_k) \nonumber \\
&&\qquad=\frac{1}{2}\sum_{\epsilon_j\in \{-1,+1\}} \prod_{j=1}^k
\epsilon_j X(1/2+\epsilon_j\alpha_j)^{-1/2} \prod_{1\leq i<j\leq
k} \zeta(1+\epsilon_i\alpha_i+\epsilon_j\alpha_j)
A(\epsilon_1\alpha_1,\ldots,\epsilon_k\alpha_k) \nonumber\\
&&\qquad \qquad\qquad \times (1+O(q^{-\tfrac{1}{2}+\epsilon})),
\nonumber
\end{eqnarray}

\end{conjecture}
\noindent where in all of the above
\begin{eqnarray}
&&A(\epsilon_1\alpha_1,\ldots,\epsilon_k\alpha_k) = \prod_p
\prod_{1\leq i<j\leq k} \left(
1-\frac{1}{p^{1+\epsilon_i\alpha_i+\epsilon_j\alpha_j}}\right)\nonumber \\
&&\qquad\qquad\times \frac{2}{\pi} \int_{0}^{\pi} \sin^2\theta
\prod_{j=1}^k \frac{e^{i\theta} \left(
1-\frac{e^{i\theta}}{p^{\frac{1}{2}+\epsilon_j\alpha_j}}\right)^{-1}
-e^{-i\theta}
\left(1-\frac{e^{-i\theta}}{p^{\frac{1}{2}+\epsilon_j\alpha_j}}
\right)^{-1}} {e^{i\theta}-e^{-i\theta}} d\theta.
\end{eqnarray}

In the case of odd orthogonal symmetry,
$\Lambda_M(s)=(1-s)(1+s)\prod_{n=1}^{N-1}(1-e^{i\theta_n}s)(1-e^{-i\theta_n}s)$
satisfies the functional equation
\begin{equation}
\Lambda_M(s)=-s^{2N} \overline{\Lambda_M}(\tfrac{1}{s})
\end{equation}
\noindent while
\begin{equation}
\mathcal{Z}_M(s)=-s^{-N}\Lambda_M(s)
\end{equation}
\noindent satisfies

\begin{equation}
\mathcal{Z}_M(s)=-\overline{\mathcal{Z}_M}(\tfrac{1}{s}),
\end{equation}
\noindent (the equivalent of (\ref{eq:functionalZf})).  The
structure of
\begin{eqnarray}
\label{eq:oddRMT}
 &&\hspace{-0.5 in}\int_{O^-(2N)}\mathcal{Z}_M (e^{-\alpha_1})\cdots
\mathcal{Z}_M(e^{-\alpha_k}) dM \nonumber \\
&&\hspace{1 in} = \sum_{\epsilon_j\in \{-1,1\} } \left(
\prod_{j=1}^k \epsilon_j e^{\epsilon_j N\alpha_j} \right)
\prod_{1\leq i<j\leq k} (1-e^{-\epsilon_i
\alpha_i-\epsilon_j\alpha_j})^{-1}.
\end{eqnarray}
\noindent parallels (in the manner described in Section
\ref{sect:compRZF}) that of Conjecture \ref{conj:oddL}.

Similarly, for even orthogonal symmetry,
$\Lambda_M(s)=\prod_{n=1}^N(1-e^{i\theta_n}s)(1-e^{-i\theta_n}s)$
satisfies the functional equation
\begin{equation}
\Lambda_M(s)=s^{2N} \overline{\Lambda_M}(\tfrac{1}{s})
\end{equation}
\noindent while
\begin{equation}
\mathcal{Z}_M(s)=s^{-N}\Lambda_M(s)
\end{equation}
\noindent satisfies
\begin{equation}
\mathcal{Z}_M(s)=\overline{\mathcal{Z}_M}(1/s),
\end{equation}
and
\begin{eqnarray}
\label{eq:evenRMT}
 &&\hspace{-0.5 in}\int_{SO(2N)}\mathcal{Z}_M (e^{-\alpha_1})\cdots
\mathcal{Z}_M(e^{-\alpha_k}) dM \nonumber \\
&&\hspace{1 in}= \sum_{\epsilon_j\in \{-1,1\} } \left(
\prod_{j=1}^k e^{\epsilon_j N\alpha_j} \right) \prod_{1\leq
i<j\leq k} (1-e^{-\epsilon_i \alpha_i-\epsilon_j\alpha_j})^{-1}
\end{eqnarray}
\noindent has the same structure of Conjecture \ref{conj:evenL}.
Clearly, due to the cancellation caused by the extra $\epsilon_j$
factors in (\ref{eq:oddRMT}), the sum of (\ref{eq:evenRMT}) and
(\ref{eq:oddRMT}) agrees with the Conjecture \ref{conj:fullL} in
the usual way.

\section*{Acknowledgements}

We are grateful to Peter Forrester for introducing NCS to the
method which was the starting point for this work.  The research
was partially supported by the American Institute of Mathematics
and a Focused Research Group grant from the National Science
Foundation. The last author was also supported by a Royal Society
Dorothy Hodgkin Fellowship.

 \pagebreak


\begin{thebibliography}{10}

\bibitem{kn:andsim95}
A.V. Andreev and B.D. Simons,
\newblock Correlators of spectral determinants in quantum chaos,
\newblock {\it Phys. Rev. Lett.}, {\bf 75}(12):2304--7, 1995.

\bibitem{kn:basfor94}
E.L. Basor and P.J. Forrester,
\newblock Formulas for the evaluation of {T}oeplitz determinants with rational
  generating functions,
\newblock {\it Mathematische Nachrichten}, {\bf 170}:5--18, 1994.

\bibitem{kn:bogkea95}
E.B. Bogomolny and J.P. Keating,
\newblock Random matrix theory and the {R}iemann zeros {I}: three- and
  four-point correlations,
\newblock {\it Nonlinearity}, {\bf 8}:1115--1131, 1995.

\bibitem{kn:bogkea96}
E.B. Bogomolny and J.P. Keating,
\newblock Random matrix theory and the {R}iemann zeros {II}:$n$-point
  correlations,
\newblock {\it Nonlinearity}, {\bf 9}:911--935, 1996.

\bibitem{kn:brezhik00}
E.~Br\'ezin and S.~Hikami,
\newblock Characteristic polynomials of random matrices,
\newblock {\it Communications in Mathematical Physics}, {\bf 214}:111--135,
  2000,
\newblock ar{X}iv:math-ph/9910005.

\bibitem{kn:bumdia02}
D.~Bump and P.~Diaconis,
\newblock Toeplitz minors,
\newblock {\it Journal of Combinatorial Theory, Series A}, {\bf 97}:252--271,
  2002.

\bibitem{kn:confar00}
J.B. Conrey and D.W. Farmer,
\newblock Mean values of {$L$}-functions and symmetry,
\newblock {\it Int. Math. Res. Notices}, {\bf 17}:883--908, 2000,
\newblock ar{X}iv:math.nt/9912107.

\bibitem{kn:cfkrs}
J.B. Conrey, D.W. Farmer, J.P. Keating, M.O. Rubinstein, and N.C.
Snaith,
\newblock Integral moments of zeta- and ${L}$-functions,
\newblock {\it preprint}, 2002,
\newblock ar{X}iv:math.nt/0206018.

\bibitem{kn:fyostr03b}
Y.V. Fyodorov and E.~Strahov,
\newblock Characteristic polynomials of random {H}ermitian matrices and
  {D}uistermaat-{H}eckman localisation on non-compact {K}${\rm \ddot{a}}$hler
  manifolds,
\newblock {\it Nucl. Phys. B}, {\bf 630}(3):453--491, 2002,
\newblock ar{X}iv:math-ph/0201045.

\bibitem{kn:fyostr03}
Y.V. Fyodorov and E.~Strahov,
\newblock An exact formula for general spectral correlation function of random
  {H}ermitian matrices,
\newblock {\it preprint}, 2002,
\newblock ar{X}iv:math-ph/0204051.

\bibitem{kn:fyostr03a}
Y.V. Fyodorov and E.~Strahov,
\newblock On correlation functions of characteristic polynomials for chiral
  {G}aussian {U}nitary {E}nsemble,
\newblock {\it Nucl. Phys. B}, {\bf 647}(3):581--597, 2002,
\newblock ar{X}iv:hep-th/0205215.

\bibitem{kn:hejhal94}
D.A. Hejhal,
\newblock On the triple correlation of zeros of the zeta function,
\newblock {\it Inter. Math. Res. Notices}, {\bf 7}:293--302, 1994.

\bibitem{kn:hughes00}
C.P. Hughes, J.P. Keating, and N.~O'Connell,
\newblock Random matrix theory and the derivative of the {R}iemann zeta
  function,
\newblock {\it Proc. R. Soc. Lond. A}, {\bf 456}:2611--2627, 2000.

\bibitem{kn:hughes01}
C.P. Hughes, J.P. Keating, and N.~O'Connell,
\newblock On the characteristic polynomial of a random unitary matrix,
\newblock {\it Commun. Math. Phys.}, {\bf 220}(2):429--451, 2001.

\bibitem{kn:katzsarnak99a}
N.M. Katz and P.~Sarnak,
\newblock {\it Random Matrices, Frobenius Eigenvalues and Monodromy},
\newblock AMS, Providence, Rhode Island, 1999.

\bibitem{kn:keasna00a}
J.P. Keating and N.C. Snaith,
\newblock Random matrix theory and $\zeta(1/2+it)$,
\newblock {\it Commun. Math. Phys.}, {\bf 214}:57--89, 2000.

\bibitem{kn:keasna00b}
J.P. Keating and N.C. Snaith,
\newblock Random matrix theory and ${L}$-functions at $s=1/2$,
\newblock {\it Commun. Math. Phys}, {\bf 214}:91--110, 2000.

\bibitem{kn:mont73}
H.L. Montgomery,
\newblock The pair correlation of the zeta function,
\newblock {\it Proc. Symp. Pure Math}, {\bf 24}:181--93, 1973.

\bibitem{kn:nonzir02}
S.~Nonnenmacher and M.~Zirnbauer,
\newblock {\it Personal communication}.

\bibitem{kn:odlyzko89}
A.M. Odlyzko,
\newblock The $10^{20}$th zero of the {R}iemann zeta function and 70 million of
  its neighbors,
\newblock {\it Preprint}, 1989.

\bibitem{kn:rub98}
M.~Rubinstein,
\newblock {\it Evidence for a Spectral Interpretation of Zeros of
  ${L}$-functions},
\newblock PhD thesis, Princeton University, 1998.

\bibitem{kn:rudsar}
Z.~Rudnick and P.~Sarnak,
\newblock Principal ${L}$-functions and random matrix theory,
\newblock {\it Duke Mathematical Journal}, {\bf 81}(2):269--322, 1996.

\bibitem{kn:weyl}
H.~Weyl,
\newblock {\it Classical Groups},
\newblock Princeton University Press, 1946.

\end{thebibliography}

\end{document}